\documentclass[a4paper,twoside,11pt]{article}
\usepackage[utf8]{inputenc}

\usepackage[english]{babel}
\usepackage{graphicx}
\usepackage{subeqnarray}

\usepackage{xcolor}
\usepackage{amsmath}
\usepackage{amssymb}
\usepackage[final]{pdfpages}
\usepackage{graphicx}
\usepackage{bbold}
\usepackage{dsfont}
\usepackage{pgfplots}
\usepackage{subfig}
\usepackage{algorithm}
\usepackage{algorithmic}
\usepackage[font=small,labelfont=bf]{caption}

\usepackage[font={footnotesize}]{caption}

\usepackage[utf8]{inputenc}
\usepackage[T1]{fontenc}
\usepackage{xcolor,amsmath,amssymb}
\usepackage{geometry}
\usepackage{graphicx}
\usepackage[sort&compress,numbers]{natbib}
\usepackage{enumerate}
\usepackage{wasysym}
\usepackage[textsize=footnotesize]{todonotes}
\usepackage{url}
\usepackage{pgfplots}

\definecolor{darkblue}{rgb}{0,0,0.6}
\definecolor{darkred}{rgb}{0.6,0,0}

\usepackage[colorlinks=true, urlcolor=black, citecolor=black, linkcolor=black, hyperfootnotes=false]{hyperref}

\usepackage{authblk}

\usepackage[bitstream-charter]{mathdesign}

\usetikzlibrary{pgfplots.groupplots}

\newcommand{\bs}{\boldsymbol}

\DeclareMathOperator{\argmax}{argmax}

\DeclareMathOperator{\Diag}{Diag}

\DeclareMathOperator{\Cor}{Cor}
\DeclareMathOperator{\Tr}{Tr}

\graphicspath{{figures/}} 

\usepackage{authblk}

\title{Endogenous Liquidity Crises}

\author[1,2]{Antoine Fosset}
\author[2,3]{Jean-Philippe Bouchaud}
\author[1,2,3]{Michael Benzaquen\footnote{Email address for correspondence: \url{michael.benzaquen@polytechnique.edu}}}
\affil[1]{Ladhyx, UMR CNRS 7646, Ecole polytechnique, 91128 Palaiseau Cedex, France}
\affil[2]{Chair of Econophysics \& Complex Systems, Ecole polytechnique, 91128 Palaiseau Cedex, France}
\affil[3]{Capital Fund Management, 23-25, Rue de l’Université 75007 Paris, France}
\date{\today \vspace{-1cm}}
\begin{document}

\maketitle

\abstract{Empirical data reveals that the liquidity flow into the order book (limit orders, cancellations and market orders) is influenced by past price changes. In particular, we show that liquidity tends to decrease with the amplitude of past volatility and price trends. Such a feedback mechanism in turn increases the volatility, possibly leading to a liquidity crisis. Accounting for such effects within a stylized order book model, we demonstrate numerically that there exists a second order phase transition between a stable regime for weak feedback to an unstable regime for strong feedback, in which liquidity crises arise with probability one. We characterize the  critical exponents, which appear to belong to a new universality class.  We then propose a simpler model for spread dynamics that maps onto a linear
Hawkes process which also exhibits liquidity crises. If relevant for the real markets, such a phase transition scenario requires the system to sit below, but very close to the instability threshold (self-organised criticality), or else that the feedback intensity is itself time dependent and occasionally visits the unstable region. An alternative scenario is provided by a class of non-linear Hawkes process that show occasional ``activated'' liquidity crises, without having to be poised at the edge
of instability.}

\tableofcontents

\section{Introduction}

Why are financial markets so prone to liquidity crises and crashes? It is now well established that market volatility is too high to be explained by fluctuations of fundamental value. In particular, a large fraction of large price jumps (say, 4-$\sigma$ events at the one minute time scale \cite{joulin2008stock}, or major daily moves \cite{Cutler, Fair}) cannot be explained by significant news. These jumps seem to be rather the result of endogenous feedback loops that lead to liquidity seizures. The memory of most spectacular ones is still vivid, such as the infamous S\&P500 flash crash of May 6th, 2010~\cite{Kyle}, or the Treasury bond flash crash of October 15th, 2014. \smallskip

These recent events have triggered a large amount of controversy, in particular in the general press, pointing fingers at electronic markets and high frequency traders. However, financial markets have always been unstable. For example on May 28th, 1962, the US stock market suffered a flash crash of severity similar to the that of May 6th, 2010~\cite{zweig2010}. This happened with good old market makers and, obviously, no HFT. Upon closer scrutiny one finds that the frequency of large price moves is remarkably stable over time, once rescaled by volatility, see e.g. \cite{bouchaud2018trades}.\footnote{Note however that the frequency of co-jumps has escalated in the past decades bearing witness of a significant increase of the level of synchronization of large price movements across assets, see \cite{calcagnile2018collective,bormetti2015modelling}.} It is found to decay as a power-law of the amplitude of the price move -- the so-called ``inverse cubic-law''~\cite{gopikrishnan1998inverse}. \smallskip

A plausible general scenario is that of destabilising feedback loops resulting in liquidity breakdown. Consider for example the classic Glosten–Milgrom model \cite{glosten} relating liquidity to adverse selection. When liquidity providers believe that the quantity of information revealed by trades exceeds some threshold, there is no longer any value of the bid–ask spread that allows them to break even -- liquidity vanishes (see e.g. \cite{bouchaud2018trades}, ch. 16). Whether real or perceived, the risk of adverse selection is detrimental to liquidity. This creates a clear amplification channel that can lead to liquidity crises.\footnote{For an alternative/complementary view on liquidity crises see also \cite{dall2019does}, where we have shown that liquidity dry outs may also be understood as the result of lag effects on latent liquidity revealing.} 
To illustrate this point, imagine that the price has recently experienced a burst of volatility. This creates anxiety for liquidity providers, who fear that some information about the future price, unbeknownst to them, is the underlying reason for the recent price changes. The consequence is an increased reluctance to provide liquidity: such liquidity providers become more likely to cancel their existing limit orders and less likely to refill the limit order book with new limit orders. Less liquidity is likely to amplify the future price moves, thereby creating an unstable feedback loop which might result in a runaway trajectory. \smallskip

The present paper attempts to capture such feedback effects both empirically and through stylised models for the dynamics of order books. In Section~\ref{section:empirical}, we empirically show that event rates in the limit order book are indeed affected by past volatility. Using tick-by-tick order book data from the EURO STOXX contract, we calibrate a generalisation of the self-exciting Hawkes processes \cite{achab2018analysis, rambaldi2017role} -- nowadays commonly used in finance but initially introduced to reproduce seismic activity. In particular, we show that market orders and cancellations tend to increase when recent price changes are large, in turn diminishing the available liquidity, much as argued above. 
We then turn to the modelling part. In Section~\ref{section:model}, to study the aggregate outcomes of such feedback in a minimal setting, we consider an extended version of the \emph{Santa Fe order book model} \cite{daniels2003quantitative,smith2003statistical,farmer2005predictive}.\footnote{The \emph{Santa Fe model} stands among the first \emph{zero intelligence} order book models reproducing some statistical properties, such as the mean bid-ask spread and mean volume profiles near the best quotes.  Note however that the model is too simple to account for volatility levels, volume profile queues far from the best, or to solve the {diffusivity puzzle}~\cite{bouchaud2018trades}.} The original model consists in a collection of $N$ queues that evolve with constant additive limit order and market order arrival probability rates, and a constant cancellation rate per existing limit order. We introduce, in a minimal fashion, the effect of interest to us by letting  past prices changes feed back into the event rates. Our numerical results strongly suggest the existence of a genuine phase transition from a stable regime to an unstable regime in which liquidity crises arise, as feedback intensity is increased. We perform a finite size scaling and determine the corresponding critical exponents. 
In Section~\ref{section:simplemodel1}  we present a simpler model, more prone to analytical treatment, setting aside the dynamics of the order book, and restricting our attention to the dynamics of the spread. We argue that phase transition scenarii can be rather generic, but require markets to sit very close to the instability threshold. Another possibility is that the feedback parameter itself is time dependent and occasionally visits the unstable phase. In Section~\ref{section:simplemodel2}, we explore an alternative scenario (activation), in which occasional liquidity crises arise without having to be poised at the edge
of instability or having a time dependent feedback parameter.
In Section~\ref{section:discussion}, we conclude.


\section{Destabilizing Feedback Effects: Empirical Analysis}
\label{section:empirical}

In this section, we provide an empirical analysis of feedback effects within order book dynamics. Consider an electronic market with three event types only: limit order deposition (LO), limit order cancellation (C) and market orders (MO). \smallskip

It is already well documented that these events strongly interact with one another. A very useful framework to describe these interactions is provided by Hawkes self-exciting point processes \cite{hawkes1971spectra}, which have already been applied to order book events in \cite{achab2018analysis, rambaldi2017role, huang2015simulating, wu2019queue, morariu2018state}. Here we want to extend these studies to account not only for activity feedback but for {\it price feedback} as well, in the spirit of the Quadratic Hawkes (Q-Hawkes) model of Blanc \emph{et al.}~\cite{blanc2017quadratic}.\footnote{See also \cite{bates2019crashes} for a recent analysis of the complex interplay between intraday volatility spikes and negative stock market jumps.}

\begin{figure}[t!]
  \centering
  \includegraphics[width=\columnwidth]{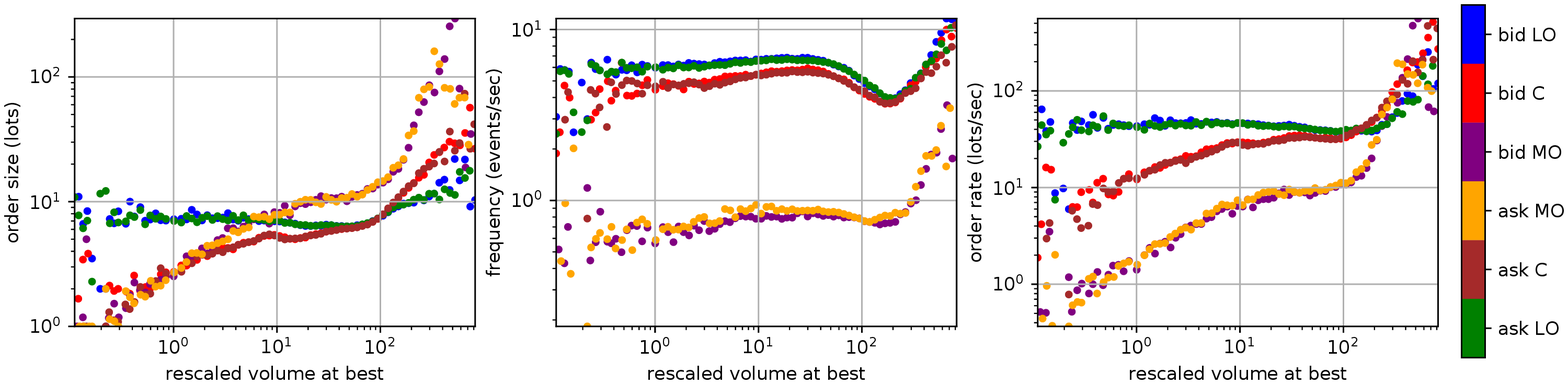}
  \caption{Average order size, frequency of events and order rate ($=$ order size $\times$ frequency) as function of volume at best rescaled by the average limit order size, on the EURO STOXX contract between $2016/09/12$ and $2017/04/28$.}
  \label{fig:rate_orderSize}
\end{figure}

\subsection{Average Event Rates}

Figure~\ref{fig:rate_orderSize} displays the average order size, frequency of events and order rate ($=$ order size $\times$ frequency) as function of the rescaled volume at best for each event type, on the EURO STOXX contract between $2016/09/12$ and $2017/04/28$. The volume at best has been rescaled by the average limit order size in the same time bin, in order to eliminate intra-day seasonalities 
In terms of time scales, we find that for EURO STOXX the average time between two events is $\tau_e =  0.03 \, s$, whereas the average time between two price changes is $\tau_p = 7 \, s$.
In addition to the expected bid-ask symmetry, Fig.~\ref{fig:rate_orderSize}(c)  reveals that the total rate of cancellations and market orders are roughly proportional to the size of the queue, whereas limit order posting does not show any appreciable dependence on the volume at best. This observation motivates the specification of the Q-Hawkes model that we calibrate below. 





\subsection{A Generalized Q-Hawkes model}

For the sake of simplicity, we focus on events (LO, C, MO) at the best quotes only, bid (b) and ask (a) (we do not distinguish between placing limit orders  at the current best or inside the spread). We therefore introduce the following six-dimensional process that counts all such events: 
$$ \boldsymbol{N}_t = \left( N_t^{\mathrm{C,b}} ,\ N_t^{\mathrm{LO,b}} , \ N_t^{\mathrm{MO,b}} ,\  N_t^{\mathrm{MO,a}} ,\  N_t^{\mathrm{LO,a}} ,\  N_t^{\mathrm{C,a}} \right)\ .$$ 
We further assume that the time dependent intensities $\bs \lambda_t$ of these six processes follow the following Q-Hawkes dynamics:
\begin{equation}\label{eq:lambda}
\boldsymbol{\lambda}_t = \boldsymbol{Q}_t  \left( \boldsymbol{\alpha}_0 + \int_0^t \boldsymbol{\phi} (t-s) \, d \boldsymbol{N}_s  + \int_0^t \boldsymbol{L} (t-s) \, d P_s 
			+  \int_0^t \int_0^t \boldsymbol{K} (t-s,t-u)  \, d P_s d P_u \right)_+ ,
\end{equation}
where $dP_s$ is the price change at time $s$ in tick units, $\boldsymbol{Q}_t = \Diag\left( Q_t^\mathrm{b}, 1,Q_t^\mathrm{b},Q_t^\mathrm{a},1,Q_t^\mathrm{a}\right)$ with $Q_t^\mathrm{b/a}$ the volume at the best bid/ask in units of average limit order size. Equation~\eqref{eq:lambda} assumes that cancellations and market orders are multiplicative while limit order event types are additive, as mentioned above. Note that all kernels $\boldsymbol{L} ,\boldsymbol{K}$ are 6-dimensional vectors and $\boldsymbol{\phi}$ a 6-dimensional matrix.\smallskip

The first term on the the RHS of Eq.~\eqref{eq:lambda} accounts for a stationary exogenous intensity $\boldsymbol{\alpha}_0$ and the second is the classical Hawkes kernel accounting for event interactions.\footnote{Whereas the Hawkes contribution is not the focus of the present paper, including it is essential to obtain a reasonable explanatory power (see below).} The third and fourth terms were introduced in \cite{blanc2017quadratic} and are new in the context of limit order book modelling. The third term is a linear feedback term from past price changes, modelling the fact that up or down price moves directly impact the rate of cancellations, market orders and limit orders. The fourth is a quadratic feedback term on the rate of order book events, which does not depend on the sign of past price changes. In \cite{blanc2017quadratic}, it was proposed to write the kernel as $\boldsymbol{K}(t-s,t-u) = \boldsymbol{K}_\mathrm{d} \psi(t-s) \delta(s-u) + \boldsymbol{K}_1 Z(t-s)Z(t-u)$, with: 
\begin{itemize}
    \item a diagonal (in time) contribution $\psi(t-s)$ with weights $\boldsymbol{K}_\mathrm{d}$, which represents the feedback of past volatility on current activity, since it can be written in terms of: 
    \[
    \Sigma^2(t) = \int_0^t \psi(t-s)  \, (d P_s)^2  ,
    \]
    \item a rank-one contribution, with weights $\boldsymbol{K}_1$, which amounts to coupling the square of past trends, as measured by: 
    \[
    R(t)=\int_0^t Z(t-s)  \, d P_s\ . 
    \]
    This is the so-called Zumbach effect: past trends, independently of their sign, lead to an increase in future activity.
\end{itemize}
In the following we will choose $\psi(s)=Z(s)=e^{-\beta s}$.\footnote{We assume for the sake of simplicity that the memory of trends is the same as that of volatility.}
One of the main empirical findings of the present study is that these two effects (volatility feedback and Zumbach effect) are indeed present and large, and capture the destabilizing feedback loop 
\[
\text{trends \& volatility} \quad  \to  \quad \text{lower liquidity}  \quad  \to  \quad \text{more trends \& volatility}
\]
as surmised in the introduction.

\begin{figure}[t!]
  \centering
  \includegraphics[width=0.5\columnwidth]{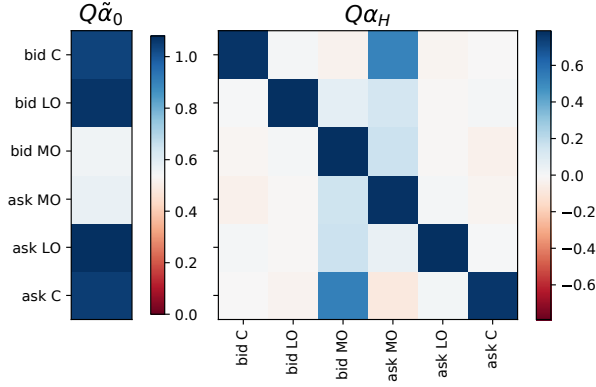}
  \caption{Fitting parameters resulting from the Hawkes non-parametric calibration on the EURO STOXX futures contract between $2016/09/12$ and $2017/04/28$. $\boldsymbol{Q} \boldsymbol{ \tilde{\alpha}}_0$ stands for the exogenous intensity from the non-parametric Hawkes fit.}
  \label{fig:calibEurostock_hawkes}
\end{figure}

\subsection{Calibration Strategy}\label{calstrat}

The Hawkes contribution $\boldsymbol{\alpha}_H = \int \boldsymbol \phi(s) ds$ has been studied in several papers in the past (see e.g. \cite{bacry2015hawkes}) and is now rather well understood. We first calibrate a Hawkes process without the price feedback term, i.e. setting $\boldsymbol{L}$ and $\boldsymbol{K}$ to zero in Eq. \eqref{eq:lambda}. We use the non-parametric technique introduced in \cite{achab2018analysis,bacry2012non}, expecting bid/ask symmetry. This means that the coefficients $\boldsymbol{Q} \boldsymbol{\tilde{\alpha}}_0$ only depend on the type of events (and not their ``sign''), and that the matrix $\boldsymbol{Q} \boldsymbol{\alpha}_H$ has a block-symmetry: the couplings of b$\to$b are equal to those of a$\to$a, and that b$\to$a is equivalent to a$\to$b. Our results are qualitatively similar to those reported in the literature~\cite{bacry2012non,bacry2015hawkes,rambaldi2017role,achab2018analysis}. The matrix structure of the norm of the Hawkes feedback kernel is shown in Fig. \ref{fig:calibEurostock_hawkes} for the EURO STOXX contract.\smallskip

Reintroducing the quadratic coupling term $\boldsymbol{K}$ leads to a much more complicated structure for the non parametric calibration problem (see \cite{blanc2017quadratic}), in particular in the present multidimensional setting. We have not yet been able to implement satisfactorily such a scheme, so we devised a simplified protocol to get some partial information on the structure of the price feedback terms. 
The idea is to capture the effect of local trends on the liquidity of the order book. Hence we define the net flux of orders at the bid $x=b$ or at the ask $x=a$ as:
\[
dJ_t^{\mathrm{x}} := dN_t^{\mathrm{LO,x}} - dN_t^{\mathrm{MO,x}} - dN_t^{\mathrm{C,x}}.
\]
From this we define the total flux and the signed flux as:
\[
dI_t^{\mathrm{b+a}} = dJ_t^{\mathrm{a}}  + dJ_t^{\mathrm{b}},  \qquad dI_t^{\mathrm{b-a}} = dJ_t^{\mathrm{b}} - dJ_t^{\mathrm{a}}. 
\]
We also introduce the forward realized flux and the forward Hawkes flux on time scale $\beta'^{-1}$: 
\[
F^\mathrm{x}_{\beta'}(t) = \int_t^{+ \infty}   e^{- \beta' (s-t)} d I_s^{\mathrm{x}} \ , \qquad
H^{\mathrm{x}}_{\beta'}(t) = \int_t^{+ \infty}   e^{- \beta' (s-t)} \lambda^{H,\mathrm{x}}_s d s  \ ,
\]
where $\mathrm{x}=(b+a,b-a)$ and $\lambda^{H,\mathrm{x}}_s$ is the expected future activity, as predicted by the Hawkes contribution.\footnote{More explicitly, $\lambda^{H,\mathrm{x}}_t := \lambda_t^{\mathrm{H,LO,x}} - \lambda_t^{\mathrm{H,MO,x}} - \lambda_t^{\mathrm{H,C,x}}$, where $\boldsymbol{\lambda}^\mathrm{H}$ is the Hawkes intensity process calibrated above. In order to speed up the computation of $F^\mathrm{x}_{\beta'}$, we approximate the non-parametric Hawkes kernels by sums of exponentials.} In the absence of other feedback mechanisms, one would expect any conditional expectation of $F^\mathrm{x}_{\beta'}(t)$ should simply be $H^{\mathrm{x}}_{\beta'}(t)$. \smallskip

\begin{figure}[t!]
  \centering
  \includegraphics[width=0.9\columnwidth]{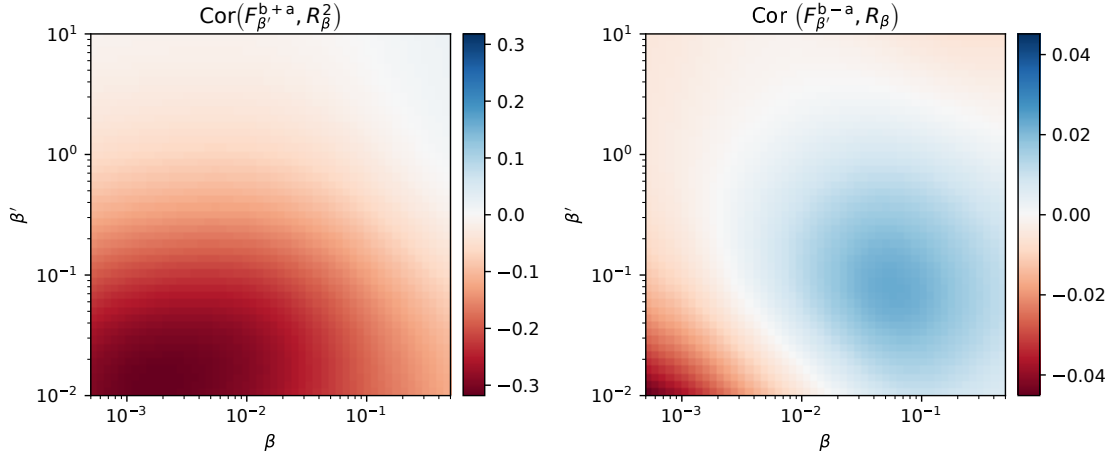}
  \caption{Correlation between the trend and the total liquidity flux (left) and signed liquidity flux (right) in the plane $\beta, \beta'$ for the EURO STOXX futures contract between $2016/09/12$ and $2017/04/28$. Note that the color scale is not the same in the left and in the right graph: the directional effect is weaker than the impact on the total (unsigned) liquidity.}
  \label{fig:correlations_trend_liquidity}
\end{figure}

This is what we test now, by considering two conditioning variables suggested by the Q-Hawkes formalism, namely past trends and past realised volatility, as measured by the following exponential moving averages:
\[
R_\beta(t) := \underbrace{\int_0^t   e^{- \beta (t-s)}  d P_s}_\text{past trend} \ , \qquad
\Sigma^2_\beta(t) := \underbrace{\int_0^t   e^{- 2\beta (t-s)}  (d P_s)^2}_\text{past volatility} .
\]
By symmetry, we expect that the conditional expectations of $F^\mathrm{b-a}_{\beta'}(t)$ 
and $F^\mathrm{b+a}_{\beta'}(t)$ write:
\begin{align}
\mathbb{E}_c [\beta' F^\mathrm{b+a}_{\beta'}(t) | R, \Sigma, H ] =& \  C_0 + 2 \beta C_1 R_{\beta}^2(t) + 2 \beta C_2 \Sigma^2_{\beta}(t) + \beta' H_{\beta'}^{\mathrm{b+a}}(t)\label{eqsC012} \\
\mathbb{E}_c [\beta' F^\mathrm{b-a}_{\beta'}(t) | R, \Sigma, H ] =& \  \sqrt{\beta} C_3 R_{\beta} + \beta' H_{\beta'}^{\mathrm{b-a}}(t) \ , \label{eqsC3}
\end{align}
i.e. the asymmetric part of the liquidity flow depends on the sign of the past trend, whereas the symmetric part of the flow depends both on the past volatility and on the past trend squared (i.e. the Zumbach effect). $C_0,C_1,C_2$ and $C_3$ are  numerical constants. The normalisation factor $\beta$ comes from the fact that $R_\beta \sim \beta^{-1/2}$ and $\Sigma^2_{\beta} \sim \beta^{-1}$. Note that the regression coefficients in front of the calibrated Hawkes contribution are fixed to unity, as they should be for consistency.

\subsection{Results}

We determine $\beta$ and $\beta'$ by looking at the maximum absolute correlations of  $F^\mathrm{b+a}_{\beta'}$ with $R_\beta^2$, see Fig. \ref{fig:correlations_trend_liquidity} and Appendix~\ref{appendix:dataanalysis}. We find $\beta = 0.001$ and $\beta '= 0.02$, corresponding to a negative correlation $\approx -0.3$, indicating that {\it trends indeed reduce liquidity}. 
Note that the correlations between $F^\mathrm{b-a}_{\beta'}$ with $R_\beta$ are one order of magnitude smaller, and in fact change sign depending on the time scales: the short time response to an up trend is adding liquidity at the ask, but the long time response is in fact removal of liquidity at the ask. This could reflect the behaviour of different actors in the market (high frequency traders/market makers vs. longer term traders). \smallskip

\begin{figure}[t!]
  \centering
  \includegraphics[width=\columnwidth]{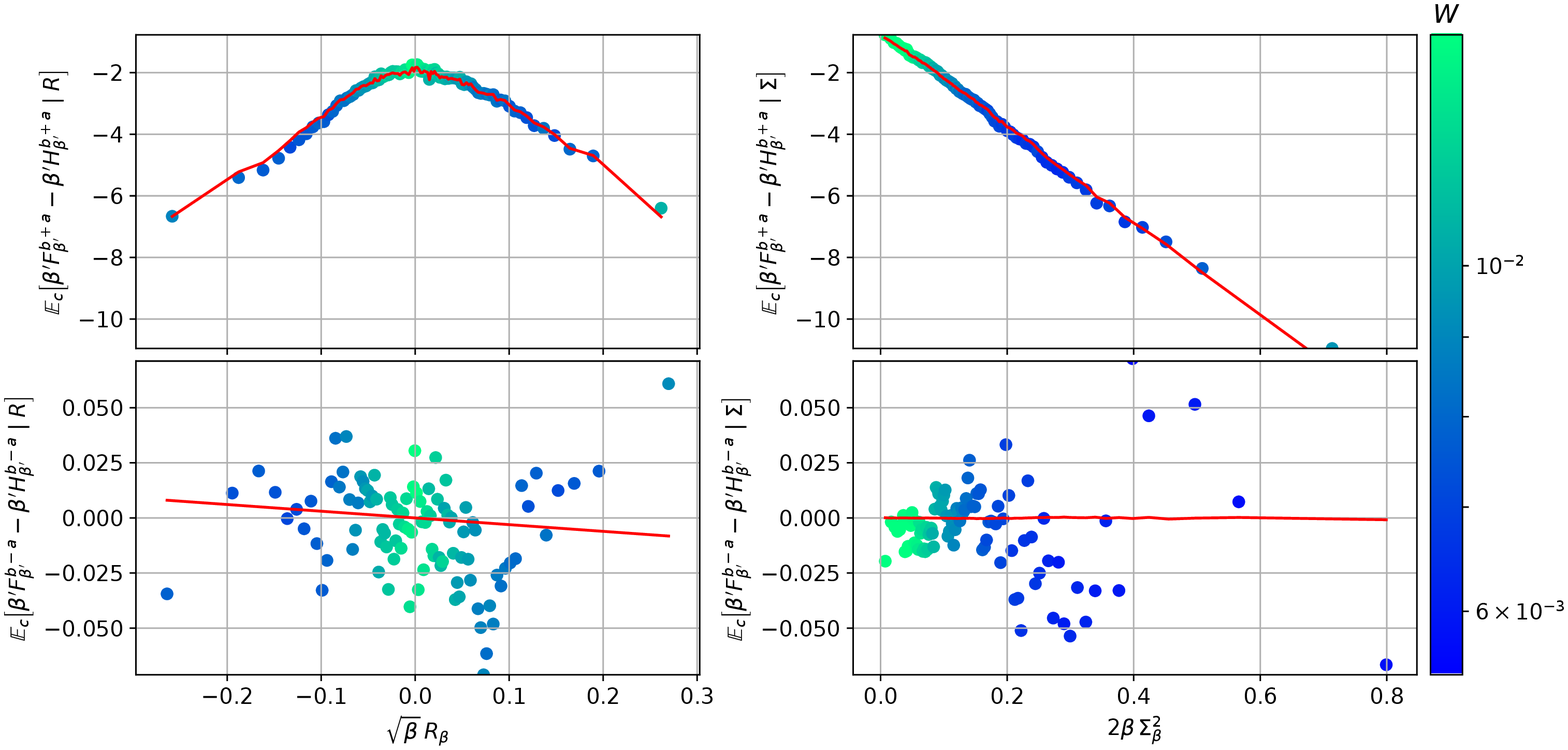}
  \caption{Regressions of the incoming flux with the trend and the volatility on the EURO STOXX futures contract between $2016/09/12$ and $2017/04/28$. The red curves correspond to the in-sample prediction of the linear regression. Each empirical point has a weight computed from the fraction of time spent in the corresponding state.}
  \label{fig:}
\end{figure}

Fixing $\beta=10^{-3}$ (i.e. trends measured over 1000 seconds, similar to the time scale found in \cite{blanc2017quadratic}) and $\beta' = 0.02$ (market response over the next 50 seconds), we find the regression coefficients $C_i$ given in Tab.~\ref{tableC} for several futures contracts, again using the period between $12/09/2016$ and $28/04/2017$. 
The quality of the regressions in the case of the EURO STOXX is illustrated in Fig. \ref{fig:} (similar plots are obtained for the BUND, BOBL and SCHATZ, not shown). We see that both the trend (Zumbach) effect, parameterised by $C_1$ and the 
volatility effect, parameterised by $C_2$, are both important to reproduce the future liquidity flow. The directional effect, measured by $C_3$, is much weaker, as indeed suggested by Fig. \ref{fig:correlations_trend_liquidity}, so we will neglect it in the following. \smallskip

The conclusions of this calibration exercise are that:
\begin{itemize}
        \item Large recent price trend and volatility indeed tend to increase the rate of market orders and cancellations and lead to a decrease in liquidity. This is the main take-away message of this section. 
        \item The quadratic feedback terms $\boldsymbol{K}$ in Eq. \eqref{eq:lambda} is the dominant effect; in the following section we will thus neglect the linear term and set $\boldsymbol{L}=0$.
\end{itemize}
The question is now whether such a quadratic feedback is enough to induce self-generated liquidity crises, i.e. an unstable feedback loop that wipes out all the volume in the order book and leads to crash. We explore this question in the next sections by analyzing different models, starting from the zero-intelligence order book model proposed by the Farmer and collaborators \cite{daniels2003quantitative,smith2003statistical,farmer2005predictive}, and ending by much simpler models amenable to some analytical predictions.  

\begin{table}[h!]
\begin{centering}
\begin{tabular}{|c|c|c|c||c|}
\hline 
     & \small $C_0 \, (10^{-2})$  & \small   $C_1$  & \small   $C_2$  & \small   $C_3$  \\ \hline
\small EUROSTOXX  & \small   $78 $  & \small   $-8.9$  & \small   $-6.7 $  & \small   $-0.03$  \\ \hline
\small BUND  & \small   $72 $  & \small   $-1.7$  & \small   $ -2.8 $  & \small   $0.16$  \\ \hline
\small BOBL  & \small   $13$  & \small   $-4.0$  & \small   $-0.29$  & \small   $0.29$  \\ \hline
\small SCHATZ  & \small   $0.42$  & \small   $-2.5$  & \small   $0.001$  & \small   $0.50$  \\
\hline  
\end{tabular}
\caption{Values of the coefficients $C_0,C_1,C_2$ for the symmetric part of the liquidity flow and $C_3$ for the antisymmetric part, as defined in Eqs.~\eqref{eqsC012} and \eqref{eqsC3}. We fixed  $\beta=10^{-3}$ and $\beta' = 2 \times 10^{-2}$.}
\label{tableC}
\end{centering}
\end{table}



\section{An Agent-Based Model for Liquidity Crises}\label{section:model}

The so-called \emph{Santa Fe model} \cite{daniels2003quantitative,smith2003statistical,farmer2005predictive} stands among the first purely stochastic order book models, where zero-intelligence agents place their orders at random (see also \cite{cont2012order,huang2015simulating}). 
It was shown that this model is able to reproduce some empirical properties of order books, such as the mean bid-ask spread and mean volume profiles near the best quotes. However the model fails to account for the empirical relation between spread and volatility (see \cite{mahadevan, wyart} and \cite{bouchaud2018trades}, Ch. 8); in fact prices are found to be strongly mean reverting, partly because of the absence of long-range correlations in the flow of market orders in the model -- see the detailed discussion of this point in \cite{toth2011anomalous, mastromatteo2014agent,bouchaud2018trades}. \smallskip

In spite of these shortcomings, the Santa Fe model is an interesting starting point for modelling order book dynamics. It consists in a collection of queues that evolve with constant additive limit and market order arrival Poisson rates, and a constant cancellation rate per existing limit order. Note that while real data is not fully consistent with additive depositions and multiplicative cancellations (see Fig. \ref{fig:rate_orderSize}), this simplifying hypothesis allows for easier analytical treatment, and leads to a well defined steady state order book where queues are neither empty nor of infinite size. \smallskip

Here, we present an extension of the Santa Fe model where the feedback of past price changes on event rates is taken into account. As suggested by the empirical results of the previous section, we only retain, for simplicity, the quadratic feedback term on cancellations, neglecting all others. We also keep the initial Santa Fe specification of an additive (rather than multiplicative) rate for market orders. Numerical simulations suggest that this brings no qualitative changes to our main conclusions, which are as follows:
\begin{enumerate}
    \item There exists a critical value of the feedback parameter $\alpha_K$ such that for $\alpha_K < \alpha^*$, an infinite size order book {\it never empties}, while for $\alpha_K > \alpha^*$ such infinite size order book {\it empties with probability 1}. 
    \item The transition appears to be of second order nature, which means that as the transition point is approached some scaling behaviour is observed. For example, the average time $\bar \tau$ needed for the liquidity crisis to appear in an infinite order book diverges as $(\alpha_K - \alpha^*)^{-\zeta}$ with $\zeta \approx 3$ when $\alpha_K \downarrow \alpha^*$. For a book of finite size $N$, this time is always finite, but diverges as $N^\eta$ with $\eta \approx 3$ when $\alpha_K=\alpha^*$. 
\end{enumerate}

\subsection{The Santa Fe Model with Feedback}

Consider a grid of prices with unit tick size, with all orders of unit size.\footnote{One could introduce a distribution of order size at the expense of extra complexity. We expect that if this distribution is broad enough, the character of the phase transition could change.} This grid is divided into three parts: the bid side $\mathcal{B}_t = \left\{p \leq b_t \right\} $, the ask side  $\mathcal{A}_t= \left\{p \geq a_t \right\}$ and the spread $ \mathcal{S}_t= \left\{b_t < p < a_t \right\} $ where $b_t$ and $a_t$ respectively denote the best bid and the best ask.  Market orders can only fall at the best bid and best ask; they do so with total rate $2\mu$, with probability $1/2$ to fall on the bid and $1/2$ to fall on the ask.\smallskip

Bid limit orders fall uniformly with rate $\lambda$ per tick size in $\mathcal{B}_t^+ = \left\{p \leq \min(b_t+1,a_t-1) \right\} $ and ask limit orders uniformly with the same rate $\lambda$ in  $\mathcal{A}_t^+= \left\{p \geq \max(a_t-1,b_t+1) \right\}$. Orders cannot be placed inside the spread at a distance higher than one tick of the best prices.\smallskip

Cancellations occur with a rate $\nu_t$ per outstanding limit order, which means that the probability that a given queue loses one order is proportional to the size of the queue. We assume that $\nu_t$ is given by a Q-Hawkes process of the type we considered in the previous section, where we retain only the Zumbach term, i.e.  
\begin{equation}
 \nu_t =  {\nu}_0 + \alpha_K \left(\int_0^t \sqrt{2 \beta}  e^{- \beta (t-s)}  d P_s \right)^2.
\end{equation}
The case $\alpha_K=0$ recovers the Santa Fe specification. Note that the dynamics of the different  price levels are independent from one another, but described by the same parameters $\lambda$ for the deposition rate and $\nu_t$ for the cancellation rate. In other words, the speed-up of cancellations when the price is trending affects {\it all} price levels.

\subsection{Numerical Simulations}

To simulate the model, we take a price grid of size $N$ ticks, and as initial condition, the equilibrium order book provided by the Santa Fe model with $\alpha_K=0$. Then, to make the system evolve one can notice that, conditioned to the past, the system follows a multidimensional, non-homogeneous Poisson process, which is well known and easy to implement. Furthermore, for computing the integral $\int_0^{t} e^{- \beta (t-s)} d P_s  = \sum_{T_n \leq t } e^{- \beta (t-T_n)} \Delta P_{T_n}$ efficiently, we use the the usual recursive formula to speed up the algorithm, see~\cite{toke2011introduction}.\smallskip

\begin{figure}[t!]
  \centering
  \includegraphics[width=\columnwidth]{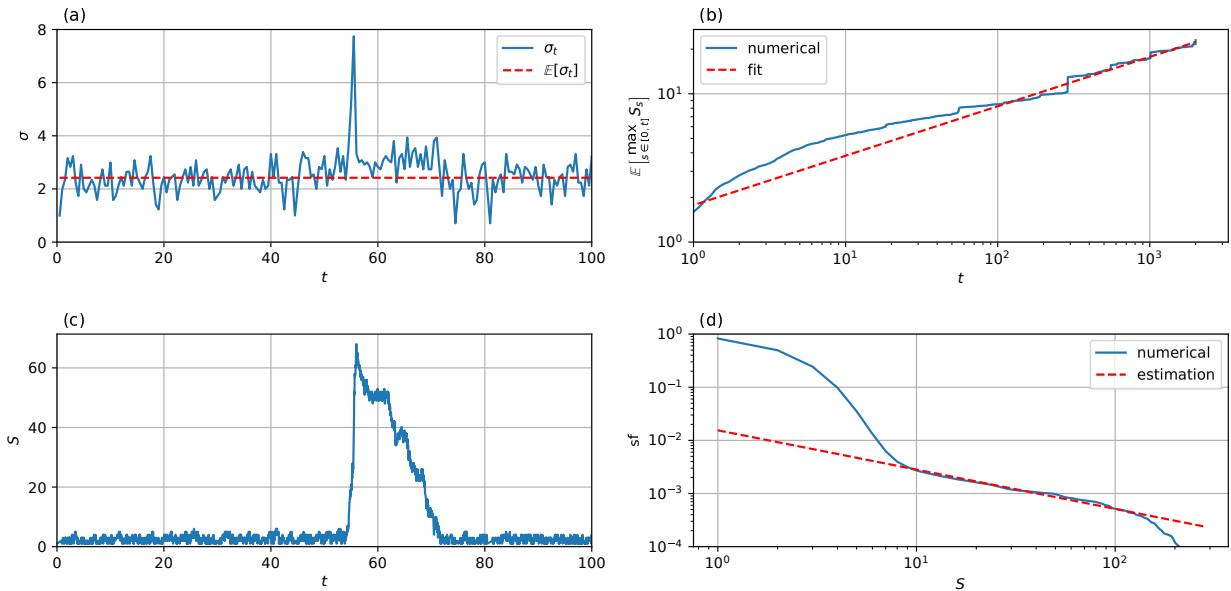}
  \caption{Properties of trajectories close to, but below the instability transition ($\lambda=10$, $\nu_0=0.5$ and $\mu=20$, and $\alpha_K=0.2$, $\beta = 1$, $T=2000$, $N=1000$). (a) One trajectory of the volatility exhibiting a cluster of high volatility. (b) Average maximum of the spread $S$ as a function of time and its fit by a power-law $t^{1/\eta}$ with $\eta=3$ for $t \geq 100$. (c) The spread trajectory corresponding to (a). (d) The spread survival function (sf), also called complementary cumulative probability distribution, decays as a power law $S^{-\kappa}$, with a cut-off that diverges as one approaches the transition $\alpha^*$. The dotted line corresponds to $\kappa=0.74$.}
  \label{fig:results_trajectories}
\end{figure}

Figure~\ref{fig:results_trajectories} displays typical results in the stable phase. Note that at some point the spread opens and triggers a cascade of cancellations that empties the order book. At some point in time denoted $\tau_c$ a liquidity crisis arises, that is here defined as the first time one side of the order book is completely empty.  


\subsection{Phase Transition and Finite Size Scaling} \label{section:fss}

\begin{figure}[t!]
  \centering
  \includegraphics[width=0.55\columnwidth]{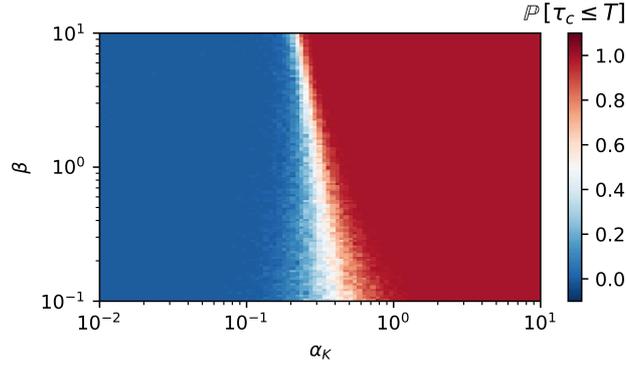}
  \caption{Stability map: Crisis probability $\mathbb P[\tau_c \leq T]$ for $T=200$, $N=280$, $\lambda = 10$, $\nu_0 = 1$ and $\mu = 20$. The blue region correspond to a stable order book, whereas the red region corresponds to liquidity crises. The crossover line $\alpha_m(\beta)$ is the white sliver between the two.}
  \label{fig:stabilityMap}
\end{figure}

Exploring the parameter space $(\alpha_K, \beta)$ reveals that for $\alpha_K \gtrsim \alpha_m(\beta)$ liquidity crises arise with high probability. Figure~\ref{fig:stabilityMap} displays the crisis probability, defined as $\mathbb P[\tau_c \leq T]$, as function of $\alpha_K$ and $\beta$ for $T \nu_0=200$ and $N=280$. As expected, large feedback intensities $\alpha_K$ lead to unstable markets. The crossover value $\alpha_m(\beta)$ decreases as $\beta$ increases, i.e. when the time scale over which trends are considered as dangerous by liquidity providers gets shorter. As expected, longer integration timescales $\beta^{-1}$ lead to more stable order book, or in other terms, longer memory is a stabilising factor.

Although suggestive, Fig.~\ref{fig:stabilityMap} cannot be used to conclude on the existence of a true phase transition in the model, between a phase where liquidity crises never happen from a phase where liquidity crises always happen, provided one waits long enough. Mathematically, the question is about the behaviour of $\mathbb P[\tau_c \leq T]$ in the double limit $N \to \infty$ and $T \to \infty$. Clearly, for finite $N$, there is always a non zero probability (perhaps very small) that the order book completely empties if one waits long enough, even when $\alpha_K=0$. Hence:
\[
\lim_{N \to \infty} \lim_{T \to \infty} \mathbb P_N[\tau_c \leq T,\alpha_K] = 1\, , \quad \forall \alpha_K\, .
\]
If one the other hand the limit $N \to \infty$ is taken first, one may be in a situation where, for a fixed value of $\beta$ 
\begin{equation}
\lim_{T \to \infty} \lim_{N \to \infty} \mathbb P_N[\tau_c \leq T,\alpha_K] = \begin{cases} 1, \quad \text{when} \quad  \alpha_K > \alpha^*, \\
0, \quad \text{when} \quad  \alpha_K < \alpha^*,
\end{cases}
\end{equation}
where $\alpha^*$ depends on the parameters of the model, in particular $\beta$. \smallskip

\begin{figure}[h]
  \centering
  \includegraphics[width =0.98\columnwidth]{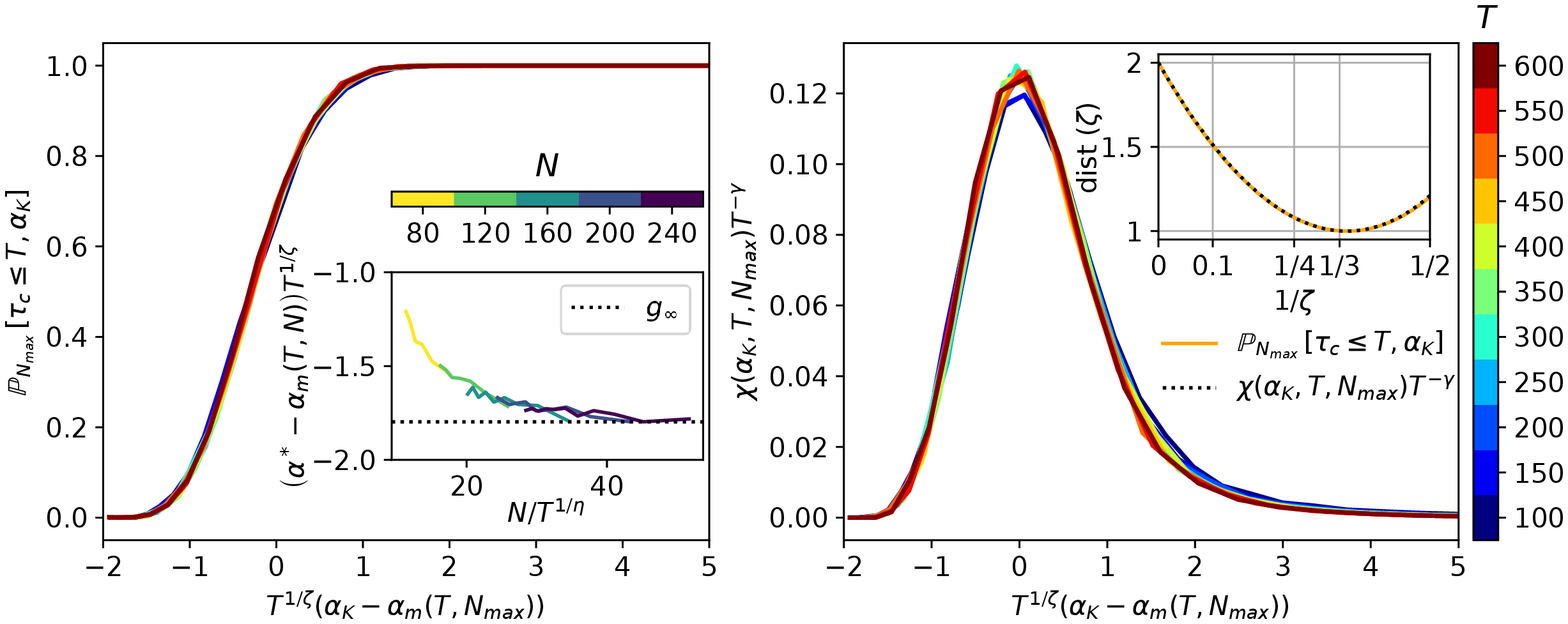}
  \caption{Left: Rescaling of $\mathbb P_N[\tau_c \leq T,\alpha_K]$ according to Eq.~\eqref{finitesize}, for $\beta=0.5$, $\lambda = 10$, $\nu_0 = 1$, $\mu = 20$, $N=N_{\max}=240$, and with $\zeta=3$ and $\eta=3$. Inset: Numerical determination of $g(v)$, also well described by our scaling hypothesis with $\eta=3$, giving $\alpha^* \approx 0.063$ and $g_\infty = \lim_{+ \infty} g \approx -1.8$.
  Right: Finite size scaling of $\chi$, with $\displaystyle \alpha_m (T,N_{\max}) := \argmax_\alpha \chi(\alpha_K, T, N_{\max})$. Inset: Average pairwise distance between the different curves as a function of $1/\zeta$, showing a minimum for $\zeta \approx 3$ for both quantities $\mathbb{P}$ and $\chi$.}
  \label{fig:finiteSizeScaling}
\end{figure}

Since numerical simulations can only be done for finite $N$ and $T$, a common strategy is to use finite size scaling to extrapolate to infinite sizes and waiting times. If a genuine, continuous phase transition occurs at some $\alpha_K=\alpha^*$, one expects the following behaviour to hold for large enough $N$ and $T$:
\begin{equation}\label{finitesize}
\mathbb P_N[\tau_c \leq T,\alpha_K] = F\left(T (\alpha_K - \alpha_m(T,N))^\zeta\right);
\qquad \alpha_m(T,N)= \alpha^* - \frac{1}{T^{1/\zeta}} g\left(\frac{N^\eta}{T}\right), 
\end{equation}
with $F(u)$ a monotonic regular function going from $0$ for $u \to -\infty$ to $1$ for $u \to + \infty$, and $g(v)$ another function that goes to a constant $g_\infty$ when $v \to \infty$ and to $+\infty$ as $v \to 0$. This scaling form has the following interpretation:
\begin{itemize}
    \item When $1 \ll T \ll N^\eta$, $\alpha_m \approx \alpha^*$. As $\alpha_K$ increases, $\mathbb P_N[\tau_c \leq T,\alpha_K]$ evolves from $0$ (no crises) to $1$ (crises) in a region of width $T^{-1/\zeta}$ around $\alpha^*$.
    \item When $T \gg N^\eta$, $\alpha_m$ becomes negative, meaning that $\mathbb P_N[\tau_c \leq T,\alpha_K]$ is close to $1$ for any $\alpha_K$ if one waits long enough.
\end{itemize}
The comparison between $T$ and $N^\eta$ has the following interpretation: for $T \ll N^\eta$, the system cannot ``feel'' the boundaries of the order book because the spread has never grown so large: $S(T) \ll N$. For $T \gg N^\eta$ on the other hand, it is highly probable that the spread $S$ has been as large as size of the order book $N$, meaning that a liquidity crisis has taken place. This suggests a direct way to measure $\eta$, from the dynamics of the spread that behaves as a power-law of time (see Fig. \ref{fig:results_trajectories}), with an exponent which should equal $1/\eta$ for consistency. This gives $\eta \approx 3$, which is compatible with the finite size scaling analysis reported in Fig. \ref{fig:finiteSizeScaling} (left inset). \smallskip

A convenient method to pin down the values of $\alpha^*$ and the exponent $\zeta$ is to study the variance of the first crisis time, defined as
\[
\chi(\alpha_K,T,N) = \mathbb{V} \left[ \min(\tau_c, T) \right] 
\]
for a fixed value of $\beta$ and different values of $\alpha_K, T$ and $N$. This quantity is expected to peak close to the phase transition, since for small $\alpha_K$, $\tau_c$ is nearly always larger than $T$ and $\chi \to 0$, whereas for large $\alpha_K$, $\tau_c$ is small and $\chi$ is also small. The finite size scaling assumption for this quantity amounts to:
\begin{equation}
   \chi (\alpha_K, T,N) = T^{\gamma} G\left(T (\alpha_K - \alpha_m(T,N))^\zeta\right),
\end{equation}
where $\alpha_m$ is given by Eq.~\eqref{finitesize} with $\alpha^* \approx 0.06$ and $G(u)$ is a humped function that goes to zero for $u \to \pm \infty$. The details and justification of this procedure to find the different exponents is described in Appendix~\ref{appendix:finitesizescaling}. We find $\gamma \approx 2$ and  $\zeta \approx \eta  \approx 3$.\footnote{Note that the value $\gamma=2$ is not unreasonable since for $\alpha_K=\alpha^*$ one expects that $\tau_c$ is larger than $T$ with some probability $p \in ]0,1[$, leading to $\mathbb{V} \left[ \min(\tau_c, T) \right] \propto T^2$.} Figure~\ref{fig:finiteSizeScaling} shows how all the different curves re-scale on top of each other when these parameters are fixed. We also show the quality of this rescaling as a function of $\zeta$ in the inset of Figure~\ref{fig:finiteSizeScaling} (right), clearly favoring the value $\zeta = 3$. \smallskip

We note that to the best of our knowledge, the numerical value of the exponents $\zeta, \eta$ do not seem to relate to an identified  phase transition. It would be very interesting to explore further the nature of this transition and (if possible) compute analytically the value of these exponents. 

\subsection{Discussion}

Although not perfect, we consider the rescaling sufficiently convincing to support our interpretation that the observed liquidity transition is a second order phase transition. This interpretation is further supported by the fact that a similar finite size scaling with the same value of the exponents $\zeta, \eta$ (but different values of $\alpha^*$) holds for different values of the time scale $\beta$ and rates $\lambda, \nu_0$ and $\mu$, and is also robust against changes in the specification of the model. This universality is a landmark of second order phase transitions. \smallskip

Although our numerical evidence for such a phase transition is satisfactory, we have not found a way to bolster our results by a rigorous mathematical analysis. Indeed, even if highly stylized, the Santa Fe model with feedback is in fact quite complex. Hence, the existence of this phase transition, and its second-order nature, can only be considered as conjectures at this stage. In order to make some progress, we have studied even simpler models, where the existence of a phase transition can be ascertained mathematically. This is what we discuss in the following sections. 

\section{A State-Dependent Hawkes Model for Spread Dynamics}
\label{section:simplemodel1}

In this section we introduce and discuss a family of simple models for which the liquidity crisis transition observed within the Santa Fe model can be analyzed in more details. This however comes at the price of setting aside the dynamics of the order book, and restrict our attention to the dynamics of the spread.

\subsection{A Simple Model}
\label{section:linlimit}
The simplest class of models consist in retaining the feedback of the spread dynamics on itself, forgetting about the price dynamics which is the main driver of the instability in the context of the Santa Fe model. Hence we also move away from the Quadratic Hawkes model we calibrated on real data in Section~\ref{section:empirical}. The destabilizing mechanism we imagine is that spread opening events are likely to lead to more spread opening events. A simplified model reinstating the price feedback mechanism is presented in Appendix \ref{appendix:price_feedback}.

We consider an order book filled with limit orders of size unity that can be cancelled or executed only at the best. As soon as a price slot is filled, no further limit order can be placed. Limit orders can thus only be placed in front of the best (inside the spread) provided the spread is open, \textit{i.e.} $S_t:= a_t - b_t \geq 2$. Note that with such simplifying hypotheses, there is no gap in the order book (apart from the spread itself) and market orders play the exact same role as cancellations. The model can thus be entirely characterized by the spread dynamics. 
We assume that the event intensities read: 
\begin{subeqnarray}\label{eq:model1}
         \displaystyle \lambda_t^+ &=& \lambda_0^+ + \alpha \int_0^t \beta e^{- \beta (t-s)} d S_s^+ \\
         \displaystyle \lambda_t^{-} &=& \mathds{1}_{\{ S_t \geq 2 \}} \lambda_0^{-} \slabel{eq:lambdamoins}
\end{subeqnarray}
where $\lambda^+$ is the intensity of events that increase the spread, \textit{i.e.} orders that are cancelled or executed by a market order and $\lambda^{-}$ is the rate of limit orders reducing $S_t$ by falling inside the spread. Only spread opening events contribute to the feedback on $\lambda^+$, i.e. $d S_t^+ := \max(d S_t,0)$. 
This highly stylized model has the advantage of being analytically tractable, while giving valuable insights on the possible phase transitions that can take place in order book models with feedback.\smallskip

For $ \alpha < 1 $, using linear Hawkes theory, one can show (see e.g.~\cite{bacry2012non}) that there exists a martingale process $M_t$ such that:
\begin{equation}\label{eq:firstmodel_spread}
    S_t = S_0 + \int_0^t \left[ \left(1 - \alpha e^{- (1-\alpha)\beta s} \right) \frac{\lambda_0^+}{1-\alpha}  - \mathds{1}_{\{ S_s \geq 2 \}} \lambda_0^{-} \right] ds + M_t\ .
\end{equation}
Introducing the parameter $\alpha_c = 1 - \lambda_0^+ / \lambda_0^{-} $ one can distinguish between the different regimes:\footnote{These results are general to any Hawkes kernel $\phi$ provided $||\phi||=\alpha<1$.}
\begin{itemize}
    \item $0 \leq \alpha < \alpha_c$ -- The system is Hawkes-stable and the spread has a stationary distribution.
    \item $\alpha_c < \alpha < \alpha^*=1$ -- The system is Hawkes-stable but the spread increases on average linearly with~$t$.
    \item $\alpha \geq \alpha^*=1$ -- The system is Hawkes-unstable, or ``explosive''.
\end{itemize}
The terminology Hawkes-(un)stable refers to the stability transition of a linear Hawkes process, that is, the transition between a regime where the intensities reach a stationary state from a regime where the number of events grows exponentially with time. 

\subsection{The Stable Regime}

\begin{figure}[t!]
  \centering
  \includegraphics[width=\columnwidth]{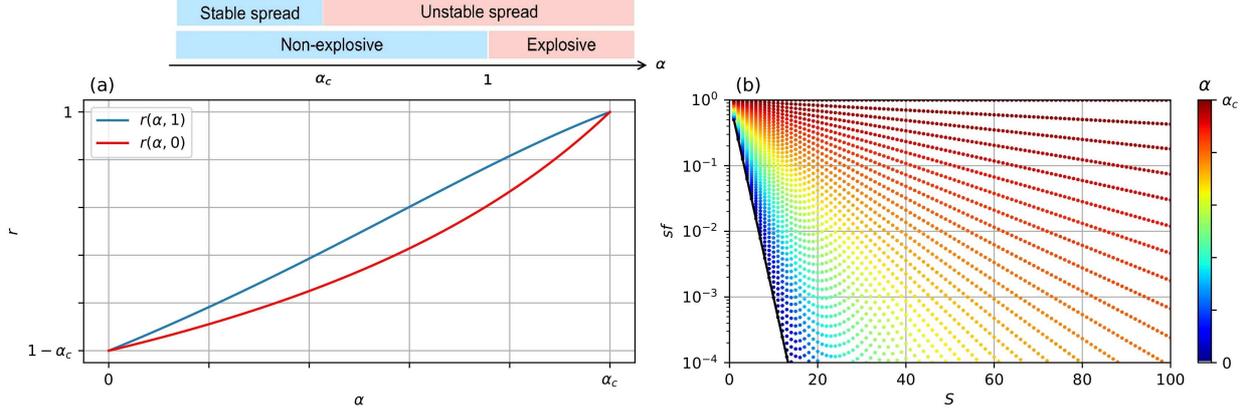}
  \caption{Properties of the spread in the linear model for $\alpha<\alpha_c$, using a set of parameters $\lambda_0^+ = 1$, $ \lambda_0^- = 0.5$ and $\beta = 1$. (a) Plot of $\alpha \mapsto r(\alpha, \beta)$ with $\beta = 1$ and the theoretical results of the limit $\beta \to 0$. (b) Log survival function (sf) of the spread for different values of $\alpha$, suggesting an exact geometrical distribution for all $\alpha$. The black curve corresponds to the theoretical equilibrium distribution when $\alpha = 0$. }
  \label{fig:spreadModel}
\end{figure}

Let us first discuss the stable regime $\alpha < \alpha_c$. In the stationary state, we can prove that the probability for the spread to be open is given by:
\[
\mathbb{P} \left[ S \geq 2 \right] = \frac{1 - \alpha_c}{1 - \alpha}\ ,
\]
which goes to $1$ as $\alpha \uparrow \alpha_c$. This result reproduces very well our numerical data. Although we have not been able to prove the result mathematically, numerical simulations also suggest that the full distribution of the spread is exactly geometric in this model:
\[
\mathbb{P} \left[ S\geq n  \right]=\frac{1 - \alpha_c}{1 - \alpha} \, (1-r)r^{n-2}\ ; \qquad n \geq 2 \ ,
\]
where $r$ depends of $\alpha$ and $\beta$, see Fig.~\ref{fig:spreadModel} (left). This result should in principle follow from the following equation that describes the evolution of the two-dimensional density function $\rho_t \big(S_t, \, X_t = \int_0^t  \beta e^{- \beta (t-s)} d S_s^+ \big)$:
\begin{eqnarray} \label{eq:fp}
\displaystyle {\partial_t \rho_t} &=& [\lambda_0^+ + \alpha (x-\beta)] \rho_t(S-1,x-\beta) \mathds{1}_{\{ S \geq 2, x \geq \beta \}} - [\lambda_0^+ + \alpha x] \rho_t(S,x) \nonumber \\
& &\displaystyle  + \lambda_0^- \rho_t(S+1,x) - \lambda_0^{-} \mathds{1}_{\{ S \geq 2 \}} \rho_t(S,x) + \beta {\partial_ x} ( x \rho_t(S,x) )\ ,
\end{eqnarray}
see Appendix~\ref{appendix:model1}. Setting the left-hand side to zero gives the stationary joint distribution of $S$ and $X$. However, we have not been able to make much analytical progress, except in the limit $\beta \to 0$ where a geometric distribution for $S$ indeed follows with $r=(1-\alpha_c)/(1-\alpha)$. Unfortunately, this approximation does not hold for $\beta \sim 1$ but works well for small $\beta$. Note that in the presence of a price feedback mechanism, the spread distribution acquires a power-law tail as we observed within the extended Santa Fe model (see Appendix \ref{appendix:price_feedback}).

\subsection{Linear Spread Growth}

In the interesting regime $\alpha_c < \alpha < \alpha^*=1$ phase, one finds that the spread grows on average linearly in time, with a drift $V$ that vanishes when $\alpha \downarrow \alpha_c$: 
\[
\lim_{t \to \infty} \frac1t \mathbb{E}[S_t] = V\ ; \qquad V:=\lambda_0^+ \frac{\alpha - \alpha_c}{(1-\alpha)(1-\alpha_c)}\ .
\] 
On top of this average drift, the spread has diffusive fluctuations with some diffusion constant $D$ defined as:
\[
D (\alpha):= \lim_{t \to \infty} \frac1t \left[\mathbb{E}[S_t^2] - \mathbb{E}[S_t]^2\right] = \lambda_0^- + \frac{\lambda_0^+}{(1 - \alpha)^3}\  .
\]
One can thus compute the probability that the spread exceeds some threshold $N$ before time $T$, corresponding to an empty book in the Santa Fe model. Using standard first passage time results for the one dimensional Brownian motion \cite{redner}, one has, for large $N$ and $T$ (and keeping the same notation as in Section~\ref{section:model}):
\begin{equation}
\label{eq:probacrisis_spreaddrift}
\mathbb P_N[\tau_c \leq T, \alpha] = \int_0^T {\rm d}u \frac{N}{\sqrt{2\pi D (\alpha) u^3}} e^{-\frac{(N+Vu)^2}{2D (\alpha) u}}\, . 
\end{equation}
While the spread will eventually exceed $N$ for large enough time, it is easy to see  that:
\[
\lim_{T \to \infty} \lim_{N \to \infty} \mathbb P_N[\tau_c \leq T,\alpha] = 0\ ,
\]
for all $\alpha < \alpha^* = 1$. In other words, the second order transition observed in the Santa Fe model with feedback is absent in the present setting. While a linear increase of the spread is interesting, it can hardly be called a liquidity ``crisis''.
Similarly the susceptibility $\chi$ can be easily computed using Eq.~\eqref{eq:probacrisis_spreaddrift} and one finds:
\begin{equation}
\label{eq:susceptibility_spreaddrift}
\chi(\alpha,T,N) = \mathbb{V} \left[ \min(\tau_c, T) \right] = \int_0^T {\rm d}u \frac{N (T-u)^2 \,  e^{-\frac{(N+Vu)^2}{2D (\alpha) u}}}{\sqrt{2\pi D (\alpha) u^3}}  - \left( \int_0^T {\rm d}u \frac{N (T-u) \, e^{-\frac{(N+Vu)^2}{2D (\alpha) u}}}{\sqrt{2\pi D (\alpha) u^3}}  \right)^2. 
\end{equation}
This gives us the same result as above:
\[
\lim_{T \to \infty} \lim_{N \to \infty} \chi(\alpha,T,N) = 0,
\]
i.e. no liquidity ``crisis''. Nevertheless, it is interesting to notice that $\chi$ can be written exactly in the scaling form that we used to analyze the Sante Fe model. Indeed, the result  of Eq.~\eqref{eq:susceptibility_spreaddrift} can be transformed into:
\[
 \chi(\alpha,T,N) = T^\gamma \mathcal{G} \left(N T^{-1/\eta}, {T}^{1/\zeta} (\alpha - \alpha_c) \right) \ ,
 \]
 with: 
 \[
 \mathcal{G} (x,y) =   x \int_0^1 \frac{{\rm d}u \, (1-u)^2}{\sqrt{2\pi D (\alpha_c) u^3}} \exp{ \left(-\frac{\left[x+u y \Lambda  \right]^2}{2D (\alpha_c) u} \right)}  - \left(x \int_0^1 \frac{{\rm d}u \,(1-u)}{\sqrt{2\pi D (\alpha_c) u^3}} \exp{ \left(-\frac{\left[x+u y \Lambda \right]^2}{2D (\alpha_c) u} \right)}  \right)^2\ ,
\]
$\Lambda:=\lambda_0^+ (1-\alpha_c)^{-2}$ and $\gamma=\eta=\zeta=2$. These exponents should  be compared with the values found numerically for the generalized Santa Fe model: $\gamma \approx 2$, $\eta \approx \zeta \approx 3$.

\subsection{The Explosive Regime}

When $\alpha > \alpha^*=1$, the model becomes Hawkes unstable, which means in the present context that the spread increases exponentially with time. Although formally the spread never diverges in finite time, in practice there is a ``liquidity crisis'' as soon as $T(\alpha - \alpha^*) \propto \log N$, i.e. when the spread reaches the boundary of the order book. This would look numerically akin to a second order phase transition with exponents $\zeta=1$ and $\eta=0$, quite far from the results reported for the Santa-Fe model.  

\subsection{A Stabilizing Mechanism}

One could expect some stabilizing mechanisms to arise when the spread becomes too large. A way to include the latter in our simple setting by substituting Eq.~\eqref{eq:lambdamoins} with:
\begin{equation}\label{eq:secondmodel_lambdamoins}
\lambda_t^- = \lambda_0^- (S_t - 1) \ ,
\end{equation}
meaning that there is an increased probability to introduce limit orders inside the spread when it is large. The model remains analytically tractable; the bottom line is that the Hawkes stable regime $\alpha_c < \alpha < 1$ disappears: our specification is indeed able to stabilise the spread in the whole region $\alpha < 1$. The Hawkes unstable regime $\alpha > 1$ of course subsists and is associated to liquidity crises in an otherwise stable market ($\alpha < 1$).  

\section{Non-Linear Hawkes Models and Metastability}
\label{section:simplemodel2}

We have studied in the previous section a simple spread dynamics model that maps onto a linear Hawkes process. In these models, the spread becomes unstable and grows linearly in time before the Hawkes process (i.e. the activity of the process) becomes itself explosive.
One can stabilize the spread dynamics, as in the last subsection above, such that sudden liquidity crises in this model are associated to the Hawkes explosive transition.\smallskip

For this picture to be correct, however, real financial markets must sit below, but very close to the Hawkes instability threshold $\alpha^*$, or else one must argue that $\alpha$ itself is time dependent, and occasionally visits the explosive region $\alpha > \alpha^*$ before decreasing back below $\alpha^*$, allowing the market to re-stabilise. 
The same remark in fact applies to the generalised Santa Fe model studied in section \ref{section:model}: if liquidity crises are indeed related to the existence of a second order phase transition, one must argue that financial markets are for some reason close to the critical point -- a phenomenon called ``self-organized criticality'' \cite{bak59self} -- or that the parameters fluctuate over time and occasionally push the system in the unstable phase.   \smallskip

Although many models in mathematical finance are tweaked such that their parameters become time dependent, we feel that this common procedure might in fact hide the inadequacy of such models. In this section, we want to explore an alternative scenario. We introduce a class of non-linear Hawkes process that show occasional liquidity crises without either being poised at the edge of instability ($\alpha \uparrow \alpha^*$) or having a time dependent feedback parameter $\alpha$.

\subsection{A Model with Quadratic Feedback}

Let us consider again the simplified framework of section~\ref{section:linlimit} and generalize the feedback on spread opening events as:
\begin{equation}\label{eq:model1bis}
\lambda_t^+ = \lambda_0^+ + \alpha X_t + \epsilon X_t^2\ ; \qquad X_t:=\int_0^t \beta e^{- \beta (t-s)} d S_s^+\, . 
\end{equation}
When $\epsilon=0$, this Hawkes process is non-explosive provided $\alpha < \alpha^* = 1$. But as soon as $\epsilon >0$, the process has a non zero probability to explode, even when $\alpha < \alpha^*$. However, interestingly, these ``liquidity crises'' only happen with a rate that is \textit{exponentially small} in $1/\epsilon$, and therefore interrupt very long periods of apparent market stability -- a phenomenon called ``metastability'' in the physics literature. This is confirmed by direct numerical simulations of the model Eq.~\eqref{eq:model1bis} in Fig.~\ref{fig:volmodel}. In the following section, we give an analytical description of this phenomenon.

\begin{figure}[t!]
  \centering
  \includegraphics[width=\columnwidth]{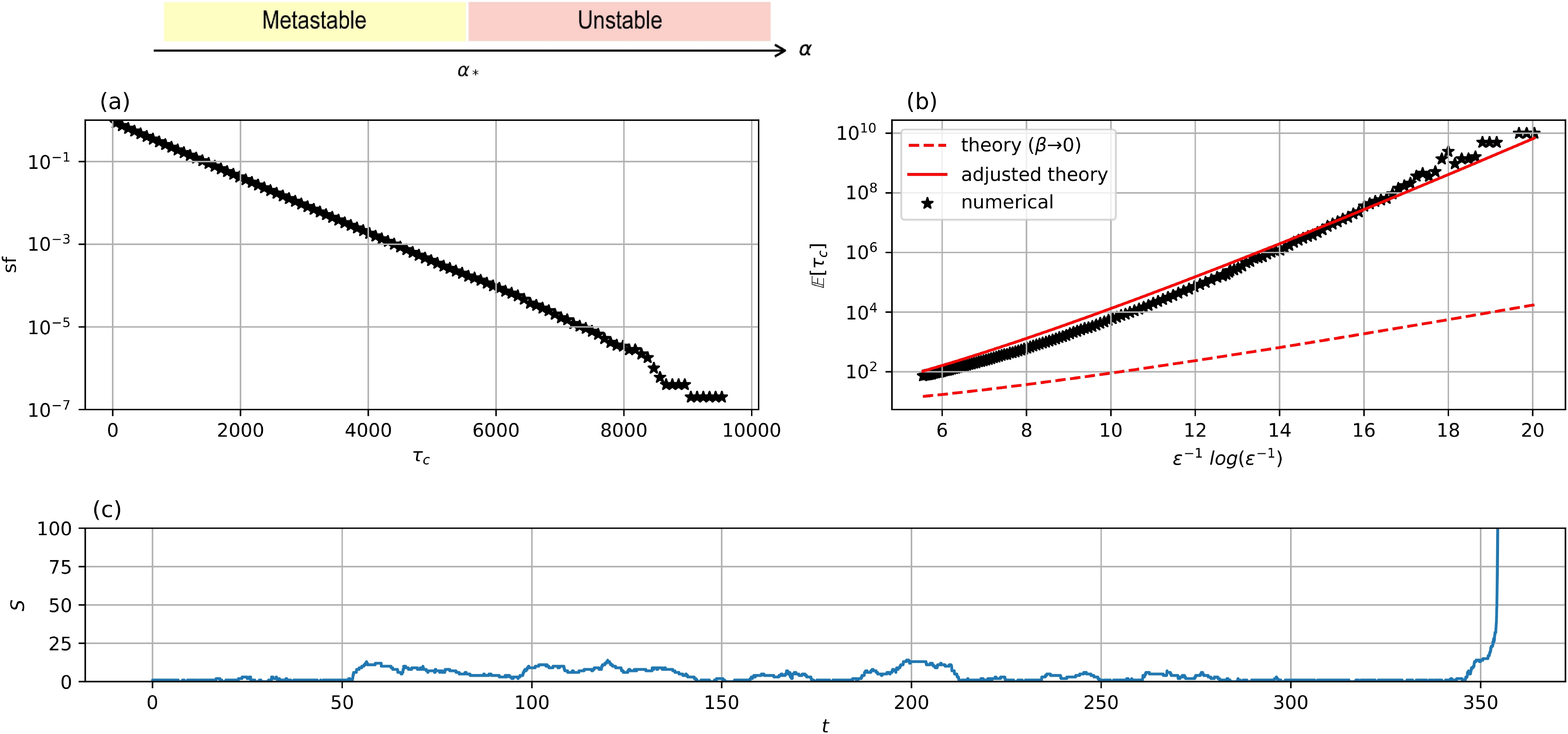}
  \caption{Properties of the time of metastability when $\alpha = 0$ and $\epsilon > 0$. (a) Survival function (sf) of the time before explosion for $\epsilon=0.2$, which is found to be exponential. (b) Evolution of the average metastability time with $\epsilon$. The dotted red curve is the continuous time prediction given by Eq.~\eqref{eq:logmeta}. The plain red curve is obtained by multiplying the term in the exponential by a empirical factor $2.5$. (c) Typical metastable trajectory. The set of parameters is the same that in (a): $\lambda_0^+ = 1$, $ \lambda_0^- = 0.5$, $\beta = 1$ and $\epsilon = 0.2$.}
  \label{fig:quad_model}
\end{figure}

\subsection{A Continuous Time Description}

\begin{figure}[t!]
  \centering
  \includegraphics[width=0.64\columnwidth]{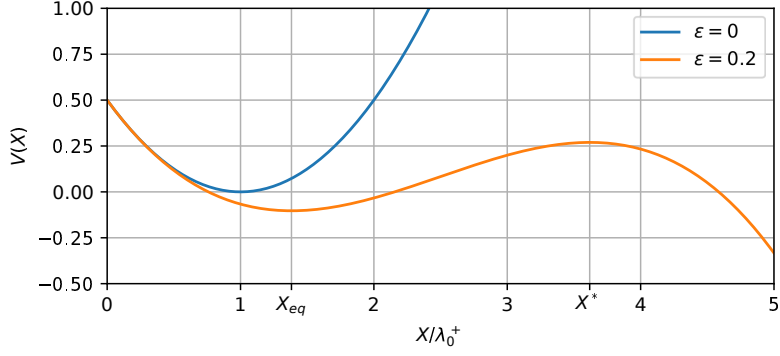}
  \caption{Plots of the potential as given by Eq.~\eqref{pot} with $\lambda_0^+ = 1$ and $ \alpha = 0$.}
  \label{fig:potential}
\end{figure}

In the ``slow'' limit $\beta \to 0$ one can write an approximate SDE for $X_t$. Start from the exact expression $dX_t = - \beta X_t \, dt + \beta dS_t^+$. When $\beta$ is small, $\lambda_t^+$ is slowly varying and one can approximate $dS_t^+$ by $\lambda_t^+ \, dt + \sqrt{\lambda_t^+} \, dB_t$, where $B_t$ is a Brownian motion (for more rigorous statements, see \cite{jaisson2015limit}). Hence: 
\begin{equation}\label{eq:metastabledynamique}
   dX_t = \beta \left[ \lambda_0^+ -  (1- \alpha) X_t  + \epsilon X_t^2\right] \, dt  + \beta \sqrt{\lambda_t^+(X_t)} \, dB_t . 
\end{equation}{}
Let us write the deterministic part of this equation as minus the derivative of some ``potential'' $\mathcal{V}(X)$, to wit:
\begin{equation}
\mathcal{V}(X) = \frac{\beta(1 - \alpha)}{2} \left(X - \frac{\lambda_0^+}{1 - \alpha}\right)^2 - \frac{\beta \epsilon}{3} X^3.
\label{pot}
\end{equation}
Such a potential is drawn for $\alpha < 1$, $\epsilon=0$ and $\alpha < 1$, $\epsilon >0$ in Fig.~\ref{fig:potential}. One sees clearly that for $\epsilon=0$ the equilibrium $X_{\text{eq}}=\lambda_0^+/(1-\alpha)$ (that corresponds to the average intensity of the Hawkes process) is stable. But as soon as $\epsilon > 0$ the potential reaches a maximum for some value $X^*$ beyond which it plunges towards $- \infty$. In the limit $\epsilon \to 0$, one finds that $X^*$ is given by:
\[
X^* \approx \frac{1-\alpha}{\epsilon}\ ,
\]
corresponding to:
\[
\mathcal{V}(X^*) \approx \frac{\beta(1 - \alpha)^3}{6 \epsilon^2}\ ,
\]
which diverges when $\epsilon \to 0$.
This picture allows one to describe the dynamics of the model for $\epsilon > 0$ in intuitive terms: for a very long time, $X_t$ will oscillate around its equilibrium value $X_{\text{eq}}$ until some rare fluctuation of the Brownian noise $dB_t$ is able to bring $X_t$ close to the top of the high barrier $X^*$. In such rare circumstances, $X_t$ escapes the stable valley and runs all the way to $+\infty$ in finite time, corresponding to a ``liquidity crisis''. \smallskip

The theory of high barrier crossing under the influence of noise is very well understood. In the present case, the final formula for the average first escape time $\tau_c$ (corresponding to the ``emptying of the book'' as in section~\ref{section:model}) is given by \cite{hanggi1990reaction, bouchaud2018trades}:
\begin{equation}\label{eq:timemetastsbility}
    \mathbb{E}[\tau_c] \approx 2 \pi \left({\frac{D(X_{\text{eq}})D(X^*)} {|\mathcal{V}''(X^*)\mathcal{V}''(X_\text{eq})|}}\right)^{1/2}  \hspace{-0.2cm}\times \exp \left( \int_{X_{\text{eq}}}^{X^*} {\rm d}x \frac{\mathcal{V}'(x)}{D(x)} \right)\, ,
\end{equation}
with $D(X) := \frac{\beta^2}{2} \left( \lambda_0^+ + \alpha X + \epsilon X^2 \right)$. The expansion to second order in $\epsilon$ gives:
\begin{equation}
    \label{eq:logmeta}
    \log \mathbb{E}[\tau_c] \underset{\epsilon \to 0}{\approx} \left\{
\begin{array}{lcc} 
\displaystyle - \frac{2}{\beta} \left[ \frac{1 - \alpha - \log \alpha }{\epsilon} + \frac{\lambda_0^+}{\alpha^2} \log \frac{1}{\epsilon} \right] - \frac{1}{2} \log \frac{1}{\epsilon} & \text{if} & \alpha > 0 \\
\displaystyle \frac{1}{\beta \epsilon} \left(\log \frac{1}{\epsilon \lambda_0^+} - 2 \right) & \text{if} & \alpha = 0 \ . \\
\end{array}
\right. 
\end{equation}
Hence, as announced, the time before a crisis is exponentially large in $\epsilon^{-1}$, with logarithmic corrections for $\alpha=0$. Another prediction of this approach is that in the limit $\epsilon \to 0$, the time-to-crisis becomes a Poisson variable with mean $\mathbb{E}[\tau_c]$, as indeed found numerically (see Fig. \ref{fig:quad_model}(a)). \smallskip

Our analytical result compares well with our numerical results in terms of the overall dependence on $\epsilon$, but the numerical prefactor inside the exponential is off by a factor $\sim 2.5$. This can be traced to the fact that our numerical simulations are in a regime where $\beta/\lambda_0 = O(1)$, whereas the theoretical analysis is done in a regime where $\beta/\lambda_0 \to 0$. (see \cite{godreche1995entropy} and \cite{bouchaud2018trades}, section 5.4, where a similar phenomenon is present).

\section{Conclusion}\label{section:discussion}

Let us summarise what we have achieved in this study. Using tick-by-tick order book data on futures contracts, we were able to show that event rates are strongly affected by past price moves. In particular, large price trend and/or volatility tends to increase the rate of market orders and
cancellations, which subsequently leads to a decrease in liquidity. This, in turn, contributes to increasing volatility, which may lead to a destabilising feedback loop and a liquidity dry-out. \smallskip

Let us stress that the work presented in this paper is relevant for both effects, explaining ``regular'' excess volatility  and understanding extreme endogenous price jumps such as flash crashes. Building on such empirical evidence, we introduced an extension of the stylised Santa Fe model which accounts for the feedback of past price changes on event rates. Numerical simulations of our model revealed the existence of a second order phase transition, and more precisely a critical value of the feedback parameter below which an infinite size order book
never empties, and above which it empties with probability one. We performed a finite size scaling analysis in order to determine the critical exponents, which does not appear to be in any of the known universality classes for 1D phase transitions.\smallskip

In order to bolster our results with analytical arguments, we then considered an even simpler model, where the existence of a phase transition can
be verified mathematically.  Setting aside the dynamics of the order book, and focusing our attention to the
dynamics of the spread, we presented a model which can be mapped onto a linear Hawkes process in which spread opening events are
likely to lead to more spread opening events. We exhibited three dynamic spread regimes as function of feedback intensity: stable, linearly increasing, and exploding spread. We argued that the second regime could be stabilised and that in such a case, and within some parameter range, a phase transition from stable to unstable spread exists, much like in the extended Santa Fe model presented before, but with major quantitative differences.\smallskip

We then pointed out that for this picture to be relevant,  real
financial markets would have to sit below, but very close to the critical point, consistent with the idea of self-organised criticality (SOC), a concept first introduced in \cite{bak59self} and  developed by many in the context of game theory \cite{challet1998minority} and financial markets \cite{giardina2003bubbles,alfi2009minimal,biondo2015modeling}. Another option would be for the feedback parameter to be
itself time dependent and occasionally visit the unstable phase. \smallskip

Finally, we presented an alternative scenario which needs no \emph{a priori} proximity to the instability threshold, nor a time dependent feedback parameter. The model is a non-linear Hawkes process for which liquidity crises are ``activated" events within a metastable phase. A continuous time description allowed us to derive the typical crisis frequency as function of the model's parameters, and show that this time can be much longer than the microscopic time of the model. It is quite likely that both effects (proximity of a transition and metastability) do actually coexist in financial markets. 

Several improvements/extensions of this work would be of interest. In particular, achieving proper calibration of the Q-Hawkes model presented in Eq.~\eqref{eq:lambda} in Section~\ref{section:empirical}, perhaps through a non-parametric procedure, would help buttress our empirical results. A deeper mathematical analysis aimed at deriving the critical exponents of the extended Santa Fe model presented in Section~\ref{section:model} would be highly valuable to ascertain the new universality class we exhibited numerically. For the sake of completeness, it would naturally also be of interest to couple our second order phase transition scenarii to a mechanism that draws the systems towards the critical point, building on ideas inspired e.g. by
the Minority Game, see \cite{challet1998minority}. Research should probably also be devoted to thinking about which empirical test could help discriminating between the second order phase transition and activation scenarii.\smallskip

An important point on which we decided not to insist too much in the present study is the effect of memory timescales, apart from the empirical section in which we thoroughly analysed the intensity of the response functions as function of forward and backward memory timescales. Indeed, as shown for example in \cite{dall2019does}, lag effects can be extremely important destabilising factors that must be taken into account. Including these lag effects within the present framework is certainly a relevant extension worth investigating. 

\section*{Acknowledgments}

We thank Jonathan Donier, Iacopo Mastromatteo, Jos\'e Moran, Mehdi Tomas, Stephen Hardiman and Mathieu Rosenbaum for fruitful discussions.
This research was conducted within the \emph{Econophysics \& Complex Systems Research Chair}, under the aegis of the Fondation du Risque, the Fondation de l’Ecole polytechnique, the Ecole polytechnique and Capital Fund Management.

\clearpage

\footnotesize

\bibliographystyle{unsrt}
\bibliography{bibs}

\begin{thebibliography}{10}

\bibitem{joulin2008stock}
Armand Joulin, Augustin Lefevre, Daniel Grunberg, and Jean-Philippe Bouchaud.
\newblock Stock price jumps: news and volume play a minor role.
\newblock {\em Wilmott Magazine}, September/October:1--7, 2008.

\bibitem{Cutler}
D~M Cutler, J~M Poterba, and L~H Summers.
\newblock What moves stock prices?
\newblock {\em Journal of Portfolio Management}, 15(3):4--12, 1989.

\bibitem{Fair}
Ray Fair.
\newblock Events that shook the market.
\newblock {\em The Journal of Business}, 75(4):713--732, 2002.

\bibitem{Kyle}
Andrei Kirilenko, Albert~S Kyle, Mehrdad Samadi, and Tugkan Tuzun.
\newblock The flash crash: High-frequency trading in an electronic market.
\newblock {\em The Journal of Finance}, 72(3):967--998, 2017.

\bibitem{zweig2010}
Jason Zweig.
\newblock Back to the future: lessons from the forgoten ‘flash crash’of
  1962.
\newblock {\em Intell Invest}, 2010.

\bibitem{bouchaud2018trades}
Jean-Philippe Bouchaud, Julius Bonart, Jonathan Donier, and Martin Gould.
\newblock {\em Trades, quotes and prices: financial markets under the
  microscope}.
\newblock Cambridge University Press, 2018.

\bibitem{calcagnile2018collective}
Lucio~Maria Calcagnile, Giacomo Bormetti, Michele Treccani, Stefano Marmi, and
  Fabrizio Lillo.
\newblock Collective synchronization and high frequency systemic instabilities
  in financial markets.
\newblock {\em Quantitative Finance}, 18(2):237--247, 2018.

\bibitem{bormetti2015modelling}
Giacomo Bormetti, Lucio~Maria Calcagnile, Michele Treccani, Fulvio Corsi,
  Stefano Marmi, and Fabrizio Lillo.
\newblock Modelling systemic price cojumps with hawkes factor models.
\newblock {\em Quantitative Finance}, 15(7):1137--1156, 2015.

\bibitem{gopikrishnan1998inverse}
Parameswaran Gopikrishnan, Martin Meyer, LA~Nunes Amaral, and H~Eugene Stanley.
\newblock Inverse cubic law for the distribution of stock price variations.
\newblock {\em The European Physical Journal B-Condensed Matter and Complex
  Systems}, 3(2):139--140, 1998.

\bibitem{glosten}
Lawrence~R Glosten and Paul~R Milgrom.
\newblock Bid, ask and transaction prices in a specialist market with
  heterogeneously informed traders.
\newblock {\em Journal of financial economics}, 14(1):71--100, 1985.

\bibitem{dall2019does}
Lorenzo Dall’Amico, Antoine Fosset, Jean-Philippe Bouchaud, and Michael
  Benzaquen.
\newblock How does latent liquidity get revealed in the limit order book?
\newblock {\em Journal of Statistical Mechanics: Theory and Experiment},
  2019(1):013404, 2019.

\bibitem{achab2018analysis}
Massil Achab, Emmanuel Bacry, Jean-Fran{\c{c}}ois Muzy, and Marcello Rambaldi.
\newblock Analysis of order book flows using a non-parametric estimation of the
  branching ratio matrix.
\newblock {\em Quantitative Finance}, 18(2):199--212, 2018.

\bibitem{rambaldi2017role}
Marcello Rambaldi, Emmanuel Bacry, and Fabrizio Lillo.
\newblock The role of volume in order book dynamics: a multivariate hawkes
  process analysis.
\newblock {\em Quantitative Finance}, 17(7):999--1020, 2017.

\bibitem{daniels2003quantitative}
Marcus~G Daniels, J~Doyne Farmer, L{\'a}szl{\'o} Gillemot, Giulia Iori, and
  Eric Smith.
\newblock Quantitative model of price diffusion and market friction based on
  trading as a mechanistic random process.
\newblock {\em Physical review letters}, 90(10):108102, 2003.

\bibitem{smith2003statistical}
Eric Smith, J~Doyne Farmer, L~Gillemot, Supriya Krishnamurthy, et~al.
\newblock Statistical theory of the continuous double auction.
\newblock {\em Quantitative finance}, 3(6):481--514, 2003.

\bibitem{farmer2005predictive}
J~Doyne Farmer, Paolo Patelli, and Ilija~I Zovko.
\newblock The predictive power of zero intelligence in financial markets.
\newblock {\em Proceedings of the National Academy of Sciences},
  102(6):2254--2259, 2005.

\bibitem{hawkes1971spectra}
Alan~G Hawkes.
\newblock Spectra of some self-exciting and mutually exciting point processes.
\newblock {\em Biometrika}, 58(1):83--90, 1971.

\bibitem{huang2015simulating}
Weibing Huang, Charles-Albert Lehalle, and Mathieu Rosenbaum.
\newblock Simulating and analyzing order book data: The queue-reactive model.
\newblock {\em Journal of the American Statistical Association},
  110(509):107--122, 2015.

\bibitem{wu2019queue}
Peng Wu, Marcello Rambaldi, Jean-Fran{\c{c}}ois Muzy, and Emmanuel Bacry.
\newblock Queue-reactive hawkes models for the order flow.
\newblock {\em arXiv preprint arXiv:1901.08938}, 2019.

\bibitem{morariu2018state}
Maxime Morariu-Patrichi and Mikko~S Pakkanen.
\newblock State-dependent hawkes processes and their application to limit order
  book modelling.
\newblock {\em arXiv preprint arXiv:1809.08060}, 2018.

\bibitem{blanc2017quadratic}
Pierre Blanc, Jonathan Donier, and J-P Bouchaud.
\newblock Quadratic hawkes processes for financial prices.
\newblock {\em Quantitative Finance}, 17(2):171--188, 2017.

\bibitem{bates2019crashes}
David~S Bates.
\newblock How crashes develop: intradaily volatility and crash evolution.
\newblock {\em The Journal of Finance}, 74(1):193--238, 2019.

\bibitem{bacry2015hawkes}
Emmanuel Bacry, Iacopo Mastromatteo, and Jean-Fran{\c{c}}ois Muzy.
\newblock Hawkes processes in finance.
\newblock {\em Market Microstructure and Liquidity}, 1(01):1550005, 2015.

\bibitem{bacry2012non}
Emmanuel Bacry, Khalil Dayri, and Jean-Fran{\c{c}}ois Muzy.
\newblock Non-parametric kernel estimation for symmetric hawkes processes.
  application to high frequency financial data.
\newblock {\em The European Physical Journal B}, 85(5):157, 2012.

\bibitem{cont2012order}
Rama Cont and Adrien De~Larrard.
\newblock Order book dynamics in liquid markets: limit theorems and diffusion
  approximations.
\newblock {\em Available at SSRN 1757861}, 2012.

\bibitem{mahadevan}
Ananth Madhavan, Matthew Richardson, and Mark Roomans.
\newblock Why do security prices change? a transaction-level analysis of nyse
  stocks.
\newblock {\em The Review of Financial Studies}, 10(4):1035--1064, 1997.

\bibitem{wyart}
Matthieu Wyart, Jean-Philippe Bouchaud, Julien Kockelkoren, Marc Potters, and
  Michele Vettorazzo.
\newblock Relation between bid--ask spread, impact and volatility in
  order-driven markets.
\newblock {\em Quantitative Finance}, 8(1):41--57, 2008.

\bibitem{toth2011anomalous}
Bence T{\'o}th, Yves Lemperiere, Cyril Deremble, Joachim De~Lataillade, Julien
  Kockelkoren, and J-P Bouchaud.
\newblock Anomalous price impact and the critical nature of liquidity in
  financial markets.
\newblock {\em Physical Review X}, 1(2):021006, 2011.

\bibitem{mastromatteo2014agent}
Iacopo Mastromatteo, Bence Toth, and Jean-Philippe Bouchaud.
\newblock Agent-based models for latent liquidity and concave price impact.
\newblock {\em Physical Review E}, 89(4):042805, 2014.

\bibitem{toke2011introduction}
Ioane~Muni Toke.
\newblock An introduction to hawkes processes with applications to finance.
\newblock {\em Lectures Notes from Ecole Centrale Paris, BNP Paribas Chair of
  Quantitative Finance}, 193, 2011.

\bibitem{redner}
Sidney Redner.
\newblock {\em A guide to first-passage processes}.
\newblock Cambridge University Press, 2001.

\bibitem{bak59self}
P~Bak, C~Tang, and K~Wiesenfeld.
\newblock Self-organized criticality: an explanation of 1/f noise, 1987.
\newblock {\em Phys. Rev. Lett}, 59:381.

\bibitem{jaisson2015limit}
Thibault Jaisson, Mathieu Rosenbaum, et~al.
\newblock Limit theorems for nearly unstable hawkes processes.
\newblock {\em The annals of applied probability}, 25(2):600--631, 2015.

\bibitem{hanggi1990reaction}
Peter H{\"a}nggi, Peter Talkner, and Michal Borkovec.
\newblock Reaction-rate theory: fifty years after kramers.
\newblock {\em Reviews of modern physics}, 62(2):251, 1990.

\bibitem{godreche1995entropy}
C~Godreche, JP~Bouchaud, and M~M{\'e}zard.
\newblock Entropy barriers and slow relaxation in some random walk models.
\newblock {\em Journal of Physics A: Mathematical and General}, 28(23):L603,
  1995.

\bibitem{challet1998minority}
Damien Challet and Yi-Cheng Zhang.
\newblock On the minority game: Analytical and numerical studies.
\newblock {\em Physica A: Statistical Mechanics and its applications},
  256(3-4):514--532, 1998.

\bibitem{giardina2003bubbles}
Irene Giardina and J-P Bouchaud.
\newblock Bubbles, crashes and intermittency in agent based market models.
\newblock {\em The European Physical Journal B-Condensed Matter and Complex
  Systems}, 31(3):421--437, 2003.

\bibitem{alfi2009minimal}
V~Alfi, Matthieu Cristelli, L~Pietronero, and A~Zaccaria.
\newblock Minimal agent based model for financial markets i.
\newblock {\em The European Physical Journal B}, 67(3):385--397, 2009.

\bibitem{biondo2015modeling}
Alessio~Emanuele Biondo, Alessandro Pluchino, and Andrea Rapisarda.
\newblock Modeling financial markets by self-organized criticality.
\newblock {\em Physical Review E}, 92(4):042814, 2015.

\bibitem{bochud2007optimal}
Thierry Bochud and Damien Challet.
\newblock Optimal approximations of power laws with exponentials: application
  to volatility models with long memory.
\newblock {\em Quantitative Finance}, 7(6):585--589, 2007.

\bibitem{mardiamultivariate}
KV~Mardia, JT~Kent, and JM~Bibby.
\newblock Multivariate analysis. 1979.
\newblock {\em Probability and mathematical statistics. Academic Press Inc}.

\end{thebibliography}

\clearpage

\appendix

\section{Appendix: Empirical Data} \label{appendix:dataanalysis}

We used tick-by-tick (or event-by-event) data for 4 futures contracts (EUROSTOXX, BUND, BOBL \& SCHATZ) over around $160$ trading days provided by CFM. We have chosen these assets because of their high liquidity and because they are all large tick (the spread is equal to its minimal value of 1 tick more than 99\% of the time). Before doing any specific inference on the data, we preprocess it in the following way:
\begin{itemize}
    \item We load data from $9 $am to $4$pm.
    \item Separate events displaying the same timestamps are shuffled within the millisecond without breaking causality.
    \item We restrict to the best queues only.
    \item We use the mid-price changes in tick units.
\end{itemize}{}
The number of events after cleaning is of the order of one million per day for the most liquid asset, and 50 000 for the least liquid. First, we perform a non-parametric Hawkes calibration that gives the parameters $\boldsymbol{Q} \tilde{\boldsymbol{\alpha}}_0$ and $\boldsymbol{Q \phi}$, as defined in Eq.~\eqref{eq:lambda}. 
Then, we turn to the contribution of the trend and the volatility. To do so, we compute $\beta' F_{\beta '}^\mathrm{b+a}$, $\beta' H_{\beta '}^{\mathrm{b+a}}$, $\beta' F_{\beta '}^\mathrm{b-a}$, $\beta' H_{\beta '}^{\mathrm{b-a}}$, $R_\beta$ and $\Sigma^2_\beta$, as defined in Section~\ref{calstrat}. 
For practical reasons, we approximate the kernels with a sum of three exponentials, in the spirit of \cite{bochud2007optimal}, which allows for a fast algorithm thanks to the recursivity of the exponentially weighted moving averages. 
 We associate a weight to each of these quantities that is the fraction of inter-event time, and bin the data in 100$\times$100$\times$100-sized windows for $ F_{\beta '}^\mathrm{b+a}$ with $  H_{\beta '}^{\mathrm{b+a}}$, $R_\beta$, $\Sigma^2_\beta$ and $  F_{\beta '}^\mathrm{b-a}$ with $  H_{\beta '}^{\mathrm{b-a}}$, $R_\beta$, $\Sigma^2_\beta$. 
 We aggregate the weights to get a weight for each bin, and perform the regressions given in Eqs.~\eqref{eqsC012} and \eqref{eqsC3} using a very standard generalised least square method \cite{mardiamultivariate}. We take the values of $\beta$ and $\beta'$ that maximise the absolute correlation $ \Cor (F_{\beta '}^{b+a}, R^2_\beta )$.

\section{Appendix: Finite Size Scaling Method} \label{appendix:finitesizescaling}

Here, we discuss the method used to do the finite size scaling in Section~\ref{section:fss}. First, let's recall the framework. The susceptibility writes:
\begin{equation}
   \chi (\alpha_K, T,N) = T^{\gamma} G\left(T (\alpha_K - \alpha_m(T,N))^\zeta\right) = T^{\gamma} \mathcal{G}\left(N T^{-1/\eta} , T^{1/\zeta} \left(\alpha_K - \alpha^* \right) \right) \ ,
\end{equation}
where the function $\mathcal{G}$ satisfies:
\begin{itemize}
    \item $\lim_{|y| \to + \infty} \mathcal{G}(x,y) = 0 $
    \item $\forall x, \quad y \mapsto  \mathcal{G}(x,y)$ has a unique maximum, denoted $y^*(x)$.
\end{itemize}

First, we determine $\gamma$. We introduce $ \alpha_m(T,N) =  \argmax_{\alpha_K} \chi (\alpha_K, T,N)$ and we assume that $\lim_{T ,N\to \infty} \alpha_m (T,N) =  \alpha^*$.
The idea is to look at $ \max_{\alpha_K} \chi (\alpha_K, T,N) = \chi (\alpha_m (T,N), T,N) $:
\begin{equation}
   \chi (\alpha_m (T,N), T,N) =  T^{\gamma} \max_y \mathcal{G} (N T^{-1/\eta} , y) \underset{N \to + \infty}{\longrightarrow}  T^{\gamma} \lim\limits_{x \to + \infty} \max_y  \mathcal{G} (x , y) \ .
\end{equation}
If $N_{max}^\eta \gg T$, then $ \chi (\alpha_m (T,N_{max}), T,N_{max}) \approx T^{\gamma} \lim\limits_{x \to + \infty} \max_y  \mathcal{G} (x , y)$ on our range of $T$. Note that we validity of such a hypothesis depends on the value of $\eta$, which we shall determine and self-consistently validate below. Then we compute the value of $\gamma\approx 2$ from a linear regression of $\log  \chi (\alpha_m (T,N_{max}), T,N_{max}) $ vs. $ \log T$.\smallskip

Then we determine $\zeta$. If $T,N_{max}$ are large enough that $N_{max}^\eta \gg  T$, then $ T^{1/\zeta} \left(\alpha_K - \alpha_m (T,N_{max}) \right) \approx T^{1/\zeta} \left(\alpha_K - \alpha^* \right)$ and $ \chi (\alpha_K, T,N_{max}) \approx T^{\gamma} \lim_{x \to + \infty} \mathcal{G} (x,T^{1/\zeta} (\alpha_K - \alpha^* )) $. So we plot $\chi (\alpha_K, T,N_{max})$ as a function of $ T^{1/\zeta} \left(\alpha_K - \alpha_m (T,N_{max}) \right)$ for different values of $T$ and $\alpha$ and we tune the exponent $\zeta$ to make all the curves collapse together, see Fig.~\ref{fig:finiteSizeScaling}. We can do this experiment numerically by minimising the distance between the curves as a function of $\zeta$. Adding the fact that we expect regular rational values we deduce the most likely exponent, $\zeta = 3$, see right inset of Fig.~\ref{fig:finiteSizeScaling}.\smallskip

Finally we compute $\alpha^*$ and $\eta$. By definition of $ \alpha_m (T,N)$, one has 
$
  T^{1/\zeta} (\alpha_m (T,N) - \alpha^*) = y^*(N T^{-1/\eta}).
$
Thus, if one plots $T^{1/\zeta} (\alpha_m (T,N)  - \alpha^*) $ as a function of $N T^{-1/\eta} $ for different values of $T$ and $N$, one should find a set of  parameters $\eta$ and $\alpha^*$ such that all the curves collapse together. This leads to $\alpha^*\approx 6.3\times 10^{-2}$ and $\eta \approx 3$, which is compatible with the direct result on the spread dynamics shown in Fig. \ref{fig:results_trajectories}, where one observes that $S(t) \sim t^{1/3}$. But since the finite size-finite time crossover should occur when $S(T) \sim N$, one finds that $T^{1/3} \sim N$, again leading to $\eta\approx 3$.
 
\section{Appendix: More on the Linear Spread Model}\label{appendix:model1}

Here, we focus on the linear case ($\epsilon = 0$), see Eq.~\eqref{eq:model3_spread}. In order to remain very general we rewrite the equation as $\lambda_t^+ = \lambda_0^+ + \left(\phi * d S^+\right)_t = \lambda_0^+ + \int_0^t \phi (t-s) d S_s^+ $. Point process theory teaches us that there exists two independent martingales $M_t^-$ and $M_t^+$ such that $ S_t^\pm = \lambda_t^\pm d t + d M_t^\pm $. One can write:
\begin{equation} 
    \lambda_t^+ = \lambda_0^+ + \left(\phi * \lambda^+\right)_t + \left(\phi * d M^+\right)_t\ .
\end{equation}
Assuming that $|| \phi || < 1$ one can define the resolvent $\phi_R = \sum_{n \geq 1} \phi^{*n}$, with $ \phi^{*(n+1)} = \phi * \phi^{*n} $. Note that $ (\delta + \phi_R) * \phi = \phi_R$ with $\delta$ the Dirac function. This enable to invert the above equation and obtain:
\begin{equation} 
    \lambda_t^+ = \left(1 + \int_0^t \phi_R (s) d s \right) \lambda_0^+ + \left(\psi * d M^+\right)_t \ .
\end{equation}
Combining the previous equations and introducing the martingale $M_t$ with $ d M_t =  d M^+_t - d M^-_t  + (\phi_R * d M^+)_t dt $, one gets:
\begin{equation}
    S_t = S_0 + \int_0^t \left[ \left(1 + \int_0^s \phi_R (u) d u \right) \lambda_0^+ - \mathds{1}_{\{ S_s \geq 2 \}} \lambda_0^{-} \right] ds + M_t\ .
\end{equation}
Equation~\eqref{eq:firstmodel_spread} is the particular case with $ \phi(t) = \alpha \beta e^{- \beta t}$. Choosing such a kernel one can derive the Fokker-Planck equation for the joint distribution of the variables $(S_t, X_t = \beta \int_0^t e^{- \beta (t-s)} d S_t^+)$ given in Eq.~\eqref{eq:fp}.
While we did not manage to solve this equation, we can compute the Laplace transform of the variable $X_t$ at equilibrium:
\begin{equation}
    \mathbb{E} \left[e^{-u X} \right] = \int_{\mathbb{R}^+} \rho_{X}^{st}(x) e^{-ux} d x = \exp \left( \int_0^u \frac{ \lambda_0^+ \left(1 - e^{-\beta v}\right)}{\alpha \left(1 - e^{-\beta v}\right) -\beta v} d v \right) \ ,
\end{equation}
from which we can get the cumulants. In particular, one has: $ \mathbb{E} [X] =  \lambda_0^+ / (1 - \alpha) $ and $\mathbb{V} [X] = \beta \lambda_0^+ / ( 2 (1 - \alpha)^2 ) $.
Interestingly, we can get the full stationary solution $\rho^{st}$ in two simple limit:
\begin{itemize}
    \item  $\alpha = 0$: $\rho^{st}(S,x) = (1-r) r^S \rho_{X}^{st}(x) $ with $r = \lambda_0^+/ \lambda_0^-$.
    \item $\beta \to 0$: $\rho^{st}(S,x) = \delta \left(x - \frac{\lambda_0^+}{1 - \alpha} \right) (1-r) r^S $ with $r = \frac{1 - \alpha_c}{1 - \alpha}$.
\end{itemize}
Note that the spread is geometricaly distributed in both cases.

\section{Appendix: A Model with Price Feedback on the Spread}\label{appendix:price_feedback}

\begin{figure}[t!]
  \centering
  \includegraphics[width=\columnwidth]{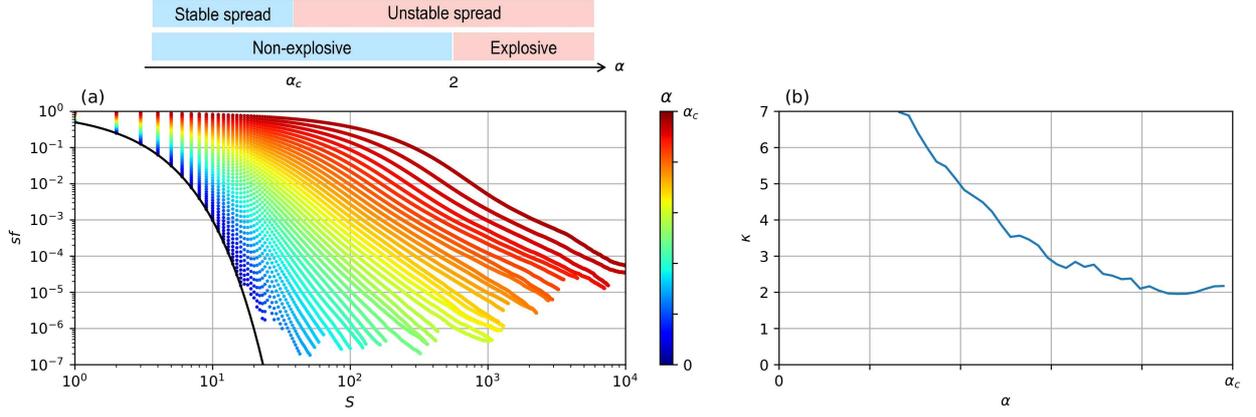}
  \caption{Properties of the spread for $\alpha<\alpha_c$ and $\beta = 1$, $\lambda_0^+ = 1$, $ \lambda_0^- = 0.5$. (a) Survival function (sf) of the spread. The black curve corresponds to the theoretical equilibrium distribution when $\alpha = 0$. For large $\alpha$, the survival function decays asymptotically as a power-law $S^{-\kappa}$. (b) Tail exponent $\kappa$ of the survival function, as a function of $\alpha$. $\kappa$ appears to saturate around $2$ when $\alpha \to \alpha_c$.}
  \label{fig:volmodel}
\end{figure}

The model presented in Section.~\ref{section:simplemodel1} displays a very simple destabilizing mechanism in the midprice reference frame, in which spread opening events trigger more spread opening events. Here, we re-introduce price dynamics to illustrate the effects of "volatility" in the spread opening mechanism, bringing the model one step closer to the empirical study presented in Section~\ref{section:empirical} and the Santa Fe model of Section~\ref{section:model}. 
In order to do so, we write the intensity of cancellations/market orders as:
\begin{subeqnarray}\label{eq:model3}
     \displaystyle \lambda_t^+ & = & \lambda_0^+ + \alpha \left( \int_0^t  \sqrt{2\beta} e^{- \beta (t-s)} d P_s\right)^2  \\
     \displaystyle \lambda_t^- & = & \lambda_0^- \mathds{1}_{\{ S_t \geq 2 \}}  \ .
\end{subeqnarray}
Each event takes place with equal probability at the bid $b_t$ or the ask $a_t$. The dynamics of the  bid and ask are thus such that
\[ 
\mathbb{E} \left[ d a_t \middle| d a_t > 0, \mathcal{F}_t \right] = -\mathbb{E} \left[ d b_t \middle| d b_t < 0, \mathcal{F}_t \right] = \lambda_t^+ d t / 2 
\]
and 
\[ 
\mathbb{E} \left[ d a_t \middle| d a_t < 0, \mathcal{F}_t \right] = - \mathbb{E} \left[ d b_t \middle| d b_t > 0, \mathcal{F}_t \right] = -\lambda_t^- d t / 2\ .
\]
Using the formalism introduced in \cite{blanc2017quadratic} we are able to solve the dynamics of the system and show that there exists a martingale $M_t$ such that:
\begin{equation}\label{eq:model3_spread}
    S_t = S_0 + \int_0^t \Big[ \left(2 - \alpha e^{- (2-\alpha ) \beta s} \right) \frac{\lambda_0^+}{2-\alpha}  - \frac{ \mathds{1}_{\{ S_s \geq 2 \}}}{2 - \alpha} \lambda_0^- \left(2 - 2 \alpha - \alpha e^{- (2-\alpha ) \beta (t-s)} \right)\Big] ds + M_t \ .
\end{equation}
Calling again $\alpha_c = 1 - \lambda_0^+ / \lambda_0^- $, one obtains the same regimes as in Section~\ref{section:linlimit} only replacing $\alpha^*=1$ by $\alpha^*=2$:
\begin{itemize}
    \item $0 \leq \alpha < \alpha_c $ -- The system is non-explosive and the spread has a stationary distribution.
    \item $\alpha_c < \alpha < 2 $ -- The system is non-explosive but the spread increases on average linearly with $t$.
    \item $2 \leq \alpha $ -- The system is explosive.
\end{itemize}
Note that this transition is similar to the Z-Hawkes transition that was presented in \cite{blanc2017quadratic}. In the $\alpha_c < \alpha < 2$ phase,  the spread grows again linearly with time:
\[ 
\mathbb{E}[S_t] \sim   Vt, \qquad V:= 2 \lambda_0^+ \frac{\alpha - \alpha_c}{(1 - \alpha_c) (2-\alpha)}.
\]
In the $\alpha < \alpha_c$ phase, one finds again $\mathbb{P} \left[ S \geq 2 \right] = ({1 - \alpha_c})/({1 - \alpha})$. Interestingly however, simulating numerically Eq.~\eqref{eq:model3} we observe that the spread distribution is asymptotically fat tailed instead of geometric (see Fig.~\ref{fig:volmodel}). Such a power law tail is also observed in our extended Santa Fe model close to the critical point.
We note that the mid-price $P_t = (a_t + b_t) / 2$ behaves like a diffusion in the two phases, with an average diffusivity $D_P$:
\[
D_P:= \lim_{t \to \infty} \frac{1}{t} \mathbb{E} \left[ \sum_{s < t} (\Delta P_s)^2 \right] =  \frac{\lambda_0^- \mathbb{P} \left[ S \geq 2 \right] + \lambda_0^+}{2 - \alpha} , 
\]
One can also show that joint the probability density function $\rho \big(t,S_t, \, X_t = \int_0^t \sqrt{2 \beta} e^{- \beta (t-s)} d p_s \big)$ now solves:
\begin{eqnarray}
  \label{eq:quad_dynamic}
 \partial_t \rho & = & \left[\lambda_0^+ + \alpha \left(x - \widehat \beta \right)^2 \right] \rho\left(t,S-1,x- \widehat \beta\right) + \left[\lambda_0^+ + \alpha\left(x +  \widehat \beta \right)^2 \right] \rho\left(t,S-1,x+ \widehat \beta\right)  \\
 & & + \;\lambda_0^- \left[ \rho\left(t,S+1,x- \widehat \beta\right) + \rho\left(t,S+1,x+ \widehat \beta\right) \right] - 2 \left[ \lambda_0^+ + \alpha x^2 + \lambda_0^- \right] \rho(t,S,x) + \beta \partial_x \left( x \rho \right) \ , \nonumber 
\end{eqnarray}
where $ \widehat \beta = \sqrt{\beta / 2}$ and with the boundary condition $ \forall S < 1, \; \rho(t,S,x)=0.$
The proof of such results uses the same techniques as in the previous appendix but is slightly more complex. First of all, there exits four martingales $ M^{-,b}_t$, $ M^{+,b}_t$, $ M^{-,a}_t$ and $ M^{+,a}_t$ such that:
\begin{equation}
  d b^\pm_t = \lambda^\pm_t d t / 2 + d M^{\pm,b}_t , \;
  d a^\pm_t = \lambda^\pm_t  d t / 2 + d M^{\pm,a}_t
\end{equation}{}
Note that we have:
\begin{eqnarray}
    d S_t & = & d a^+_t +  d b^+_t -  d a^-_t +  d b^-_t \nonumber \\
    d P_t & = & \left(d a^+_t - d a^-_t + d b^-_t - d b^+_t \right) / 2  \\
    d [P]_t & = & \left( d P_t \right)^2 =  \left(d a^+_t + d a^-_t + d b^-_t + d b^+_t \right) / 4 \ .\nonumber
\end{eqnarray}
We then use Eq.~\eqref{eq:model3} in a more general framework:
\begin{equation}
    \lambda^+_t = \lambda_0^+ + \int_0^t \int_0^t K(t-s,t-u) d P_s d P_u\ ,
\end{equation}
 where $ K$ is symmetric. Calling $\alpha = \Tr{K} = \int_0^{+\infty} K(t,t) d t$, one can rewrite: 
 \begin{equation}
     \lambda^+ = \lambda_0^+ +  \int_0^t K(t-s,t-s) d [P]_s + M^P_t = \lambda_0^+ + \left( \phi * (\lambda^+ + \lambda^-) \right)_t + \frac{1}{2} \left(\phi * (M^+ + M^-) \right)_t + M^P_t\ ,
 \end{equation}
 where $\phi(t) = K(t,t)/2 $, $ M^\pm_t = M^{\pm,a}_t + M^{\pm,b}_t$ and $M^P_t = \int_0^t \left( \int_0^{s-} K(t-s,t-u) d P_u \right) d P_s$, that is a martingale. Introducing the resolvent $\phi_R = \sum_{n\geq1} \phi^{*n}$ and the martingale: 
 \[
 d M_t = \left[  \left( \phi_R*(M^+ + M^-)\right)_t / 2 + \left( (\delta + \phi_R) * M^P \right)_t\right] d t,
 \]
 we solve the equation:
 \begin{equation}
     \lambda^+_t = \left(1 + \int_0^t \phi_R(s) d s \right) \lambda_0^+ +  \left(\phi_R * \lambda^- \right)_t + d M_t \ ,
 \end{equation}{}
and deduce the dynamics of the spread:
 \begin{equation}
     S_t = S_0  + \int_0^t \left[ \lambda_0^+ \left(1 + \int_0^s \phi_R(u) d u \right) - \lambda_0^- \mathds{1}_{\left\{ S_s \geq 2\right\}} \left(1 - \int_0^{t-s} \phi_R(u) d u \right) \right] d s + M_t \ .
 \end{equation}{}
This gives the condition of stability: if $ \alpha <  \alpha_c $ then $\mathbb P [S \geq 2] = (1 - \alpha_c) / (1 - \alpha) $. Then we get the diffusivity of the price:
\[
\lim_{t \to + \infty} \frac{1}{t} \mathbb{E} [ [P]_t ] = \lim_{t \to + \infty} \frac{1}{2} \left( \mathbb{E} [ \lambda^+_t  + \lambda^-_t]\right) = \frac{\lambda^+_0 + \lambda^-_0 \mathbb{P} [ S \geq 2 ]}{2 (1 - ||\phi||)}
\]
One can check that $||\phi|| = \alpha / 2 $.

\end{document}